\documentstyle[psfig]{mn}
\begin{document}
\hsize=6truein

\renewcommand{\thefootnote}{\fnsymbol{footnote}}

\title[]{ {\it ROSAT} PSPC observations of 36 high-luminosity clusters
of galaxies: constraints on the gas fraction}

\author[]
{\parbox[]{6.in} {S. Ettori and A.C. Fabian \\
\footnotesize
Institute of Astronomy, Madingley Road, Cambridge CB3 0HA \\
}}                                            
\date{ACCEPTED, MNRAS (1999)}
\maketitle

\begin{abstract} We present a detailed and homogeneous analysis of the
{\it ROSAT} PSPC surface brightness profiles of 36 clusters of galaxies
with high X-ray luminosity ($L_{\rm X} \ga 10^{45}$ erg s$^{-1}$) and
redshifts between 0.05 and 0.44. Using recent {\it ASCA} estimates of the
temperature of the gas for most of the clusters in the
sample, we apply both the
deprojection technique and model fitting to the surface brightness
profiles to constrain the gas and dark matter distributions under the
assumption that the gas is both isothermal and hydrostatic.

Applying robust estimators, we find that the gas fraction within $r_{500}$
of the clusters in our sample has a distribution centred on $f_{\rm
gas}(r_{500}) = 0.168 h_{50}^{-1.5}$. The gas fraction ranges from 0.101
to 0.245 at the 95 per cent confidence level. The values of $f_{\rm
gas}$ show highly significant variations between individual clusters,
which may be explained if the dark matter has a
significant baryonic component.
Within a cluster, the average radial dependence of the gas mass fraction
increases outward as $r^s$, with $s \sim 0.20$. 
Combining these results with those of primordial nucleosynthesis
calculations and the current estimate of $H_0$, the above central location
implies $\Omega_{\rm 0, m} \la 0.56$ at the 95 per cent confidence
level.
This upper limit decreases to 0.34 if we take the highest significant
estimates for $f_{\rm gas}$.

A significant decrease in cluster gas fraction with redshift from 
the local value, $f_{\rm gas, 0}$, of 0.21, found assuming $\Omega_{\rm 0, m} 
=1$, is also reduced if $\Omega_{\rm 0, m}$ is low.

\end{abstract}


\section{INTRODUCTION} 
 
The physics of the formation of clusters of galaxies depends upon the  
cosmological parameters, $\Omega_{\rm 0, m}$ and $\Omega_{\rm b}$,
that describe the observed total matter density and its baryonic
contribution, respectively. 
Assuming that clusters maintain the same ratio $\Omega_{\rm b}/
\Omega_{\rm 0, m}$ as the rest of the Universe, a measure of the
cluster baryon fraction can be compared with calculations from cosmic
nucleosynthesis considerations of the abundance of the light elements
(e.g. D, $^3$He, $^4$He, $^7$Li) to give a direct constraint on
$\Omega_{\rm 0, m}$.

In recent years, White et al. (1993), White \& Fabian (1995), David, Jones
\& Forman (1995) and others have discussed this issue first for the Coma
cluster and then for samples of clusters, highlighting the necessity
of a low density Universe in order to reconcile the baryon fraction of the
total cluster mass with the primordial $\Omega_{\rm b} \sim 0.05$.  This
``Baryon Catastrophe" for a flat Universe persists on supercluster scales;
Fabian (1991) and Ettori, Fabian \& White (1997) estimate a 15 per cent
gas contribution to the total mass of the Shapley Supercluster, over a
region of 30 Mpc in radius. 

The main baryonic component of the richest clusters is the intracluster
gas, whereas galaxies contribute less than about 4 per cent with respect
to the total gravitating mass (White et al. 1993, Fukugita, Hogan \&
Peebles 1998).
Apart from a brief discussion in Section~5, we will not consider further 
any other contribution to the total baryon budget in clusters, such as
baryonic dark matter or cool gas, for which the uncertainties are still
large (cf. Fukugita et al. 1998). Thus, the gas fraction provides a lower 
limit on the total baryon fraction in clusters.

The information on the gas and total mass distributions is inferred from
spectral and spatial analyses in the X-ray waveband. 
To properly know the gas density, we need to deproject the observed 
surface brightness profile, which is simply the projection on the sky of
the (mostly) bremsstrahlung\footnote[1]{We adopt a {\it MEKAL}
model for our analysis as described in Sect.~3} emissivity, $\epsilon \propto$
(cluster plasma temperature, $T_{\rm gas}$)$^{1/2} \times$ (gas density,
$\rho_{\rm gas}$)$^2$ (for $T_{\rm gas} \ga 3\times 10^7$ K; see e.g.
Rybicki \& Lightman 1979), i.e.:
\begin{equation}
S(b) = \int_{b^2}^{\infty} \frac{\epsilon dr^2}{\sqrt{r^2 - b^2}}.
\label{fgas:eq1}
\end{equation}

Due to the small dependence of $S(b)$ on the temperature (in particular in
the {\it ROSAT} waveband), $\rho_{\rm gas}$ is well constrained from
eqn.~\ref{fgas:eq1}.
Assuming that the hydrostatic equilibrium holds in the cluster regions
examined, we can write:
\begin{equation}
\frac{1}{\rho_{\rm gas}}\frac{d P_{\rm gas}}{dr} =
-\frac{d\phi}{dr} = -\frac{G M_{\rm tot}(r)}{r^2},
\label{fgas:eq2}
\end{equation}
where $G$ is the gravitational constant, and the gas pressure, $P_{\rm
gas}$, is calculated through the perfect gas law, $P_{\rm gas} = \rho_{\rm
gas} kT_{\rm gas} / (\mu m_{\rm p})$ (the mean molecular weight, $\mu$, is
0.6 in atomic mass unit). 
At the present, there are two unknown quantities:
the temperature profile (for sake of simplicity we assume the gas to be
isothermal, but see Sect.~4.3) and the dark matter distribution. Fixing
one of these allows us to solve the
differential equation for the other one. In particular, according to the
different cases that we discuss in Sect.~3, we adopt the dark matter
density profile found in N-body simulations and the best
available estimate of the intracluster
temperature, $T_{\rm gas}$.

In this paper, we present the analysis of {\it ROSAT} Position
Sensitive Proportional Counter (PSPC) surface brightness profiles of 36
clusters of galaxies, with X-ray luminosity greater than $10^{45}$ erg
s$^{-1}$ and redshift in the range 0.05-0.44.
These physical characteristics allow them to be well covered by the PSPC 
field of view, with the surface brightness profile extending to
about the virial radius.

Due to the energy-limited range of the PSPC (0.1--2.4 keV), we use
observations of $T_{\rm gas}$ from recent published work on {\it
ASCA} data (0.5--10 keV).

The paper is organized as follows.
In Section~2, we present the cluster sample and data reduction methods. In
Section~3, we obtain constraints on the cluster dark matter after
comparing the deprojection analysis with a straightforward fitting
approach, using both a $\beta$-model and a gas profile obtained through
the Navarro-Frenk-White dark matter profile. In Section~4, the value and
the distribution of the gas fraction, $f_{\rm gas}$, are discussed.  The
constraints that we place on $\Omega_{\rm 0, m}$ using the best estimate
of $f_{\rm gas}$, and the primordial nucleosynthesis results, are
presented in Section~5.  We summarize our main results in Section~6. 

\section{THE SAMPLE} 

\begin{figure*}
\psfig{figure=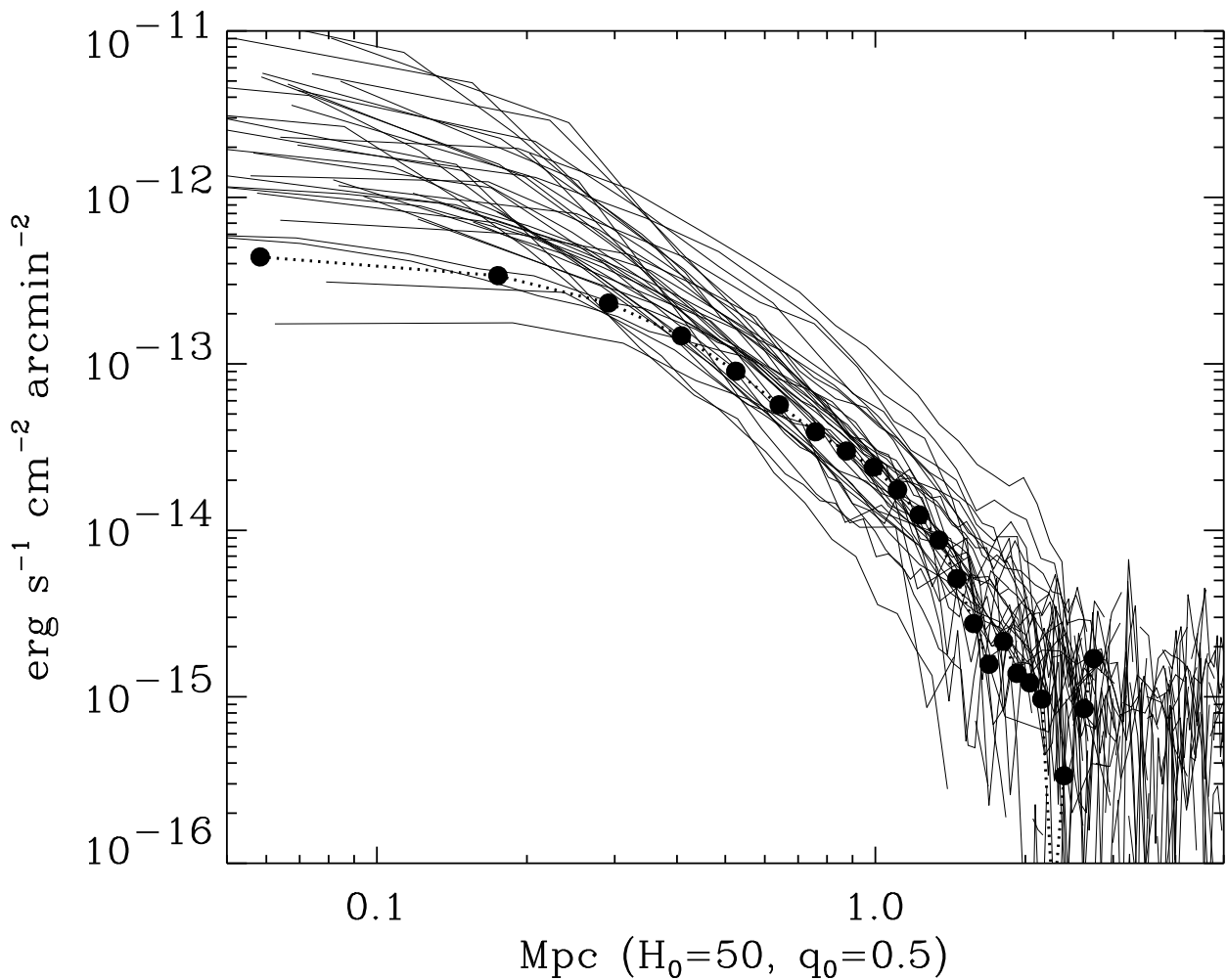,width=\textwidth,angle=90}
\caption{The surface brightness profiles of the clusters in our sample 
in the 0.5-2 keV band, converted to the flux at the rest-frame energy
band and 
corrected for the cosmological dimming by a factor $(1+z)^4$, are here 
compared with the Coma profile (dots connected by dotted line). 
} \label{fgas1} \end{figure*}


\begin{table*}   \caption{ROSAT observation summary. Reference: [1] Allen 
\& Fabian (1997); [2] B\"ohringer et al. (1998); [3] David et al. (1993);
[4] Fukazawa et al. (1998); [5] Markevitch (1998); [6] from 
$T_{\rm X}-\sigma_{\rm opt}$ relation in White, Jones \& Forman 
(1997) and $\sigma_{\rm opt} = 1307\pm 100$ km s$^{-1}$ (Girardi et al.  
1997).
{\it Note:} A2142 = rp800233 +wp800551n00 +wp800096 +wp150084. 
$R_{\rm out}$ is in Mpc, `bkg' is in units of $10^{-4}$ count s$^{-1}$
arcmin$^{-2}$.
$^{\dagger}$ This value (in units of 10$^{20}$ atoms cm$^{-2}$) is the average
of the HI column density, calculated within 1 degree and weighted with respect
to the distance from the quoted coordinates (from the map by Dickey \& 
Lockman 1990). }
\begin{tabular}{| l@{\hspace{.7em}} l@{\hspace{.7em}} r@{\hspace{.7em}} 
c@{\hspace{.7em}} c@{\hspace{.7em}} c@{\hspace{.7em}} r@{\hspace{.7em}} 
r@{\hspace{.7em}} c@{\hspace{.7em}} c@{\hspace{.7em}} c@{\hspace{.7em}} |} 
\hline
cluster & Seq. Id. & Exp. (ks) & $z$ & $\alpha_{2000}$ &
$\delta_{2000}$ & N$^{\dagger}_{\rm H}$ & $T_{\rm gas}$ (keV) & bin &
$R_{\rm out}$ & bkg \\  \hline
\multicolumn{11}{c}{  }\\
A401 & rp800235n00 &  6.6 & 0.0748 & 02$^h$ 58$^m$ 56\fs1 & $+13^{\circ}
34' 55''$ &10.5 & $8.0\pm0.2$ [5] & 30\arcsec-57.5 kpc & 2.10 & 2.18 \\
A478 & rp800193n00 & 21.8 & 0.0881 & 04 13 23.9 & +10 28 04 &15.1 &
$8.1\pm0.7$ [1] & 60-132.5 & 2.05 & 1.86 \\
A483 & rp800089n00 & 9.7 & 0.2800 & 04 15 54.8 & $-11$ 32 17 & 3.9 &
$8.7\pm2.0$ [3] & 30-158.2 & 1.19 & 5.53 \\
A520 & rp800480n00 &  4.6 & 0.2030 & 04 54 08.2 & +02 55 42 & 7.8 &
$8.3\pm0.5$ [1] & 30-128.0 & 1.86 & 2.43 \\
A545 & rp800523n00 & 13.2 & 0.1530 & 05 32 24.0 & $-11$ 32 15 &11.5 &
$5.5\pm6.2$ [3] & 90-311.8 & 2.34 & 1.93 \\
A586 & rp800348n00 &  2.9 & 0.1710 & 07 32 18.6 & +31 38 19 & 5.2 &
$10.7\pm6.3$ [1] & 30-113.0 & 1.19 & 2.25 \\
A644 & rp800379n00 &  8.7 & 0.0704 & 08 17 24.5 & $-07$ 30 29 & 6.8 &
$8.1\pm0.5$ [5] & 90-163.5 & 3.02 & 1.99 \\
A665 & rp800022n00 & 34.1 & 0.1816 & 08 30 56.3 & +65 50 52 & 4.2 &
$9.0\pm0.4$ [1] & 60-236.3 & 2.24 & 3.54 \\
A1068 & rp800410n00 & 9.5 & 0.1386 & 10 40 42.6 & +39 57 22 & 1.0 &
$5.5\pm0.9$ [1] & 30-96.3 & 1.20 & 2.27 \\
A1413 & rp800183n00 & 7.1 & 0.1427 & 11 55 17.6 & +23 24 28 & 2.2 &
$8.5\pm0.8$ [1] & 30-98.5 & 1.53 & 3.05 \\
A1651 & wp800353 & 7.1 & 0.0825 & 12 59 20.6 & $-04$ 11 30 & 1.8
&$6.1\pm0.2$ [5] & 90-187.8 & 2.91 & 3.01 \\
A1689 & rp800248n00 &12.9 & 0.1810 & 13 11 28.7 & $-01$ 20 17 & 1.8 &
$10.0\pm0.7$ [1] & 30-117.9 & 1.59 & 3.06 \\
A1763 & rp800252n00 &12.2 & 0.1870 & 13 35 17.0 & +41 00 07 & 0.9 &
$9.7\pm0.4$ [1] & 60-241.4 & 2.05 & 2.72 \\
A1795 & rp800105n00 &34.4 & 0.0621 & 13 48 51.3 & +26 35 46 & 1.2 &
$5.9\pm0.2$ [1] & 30-48.7 & 1.49 & 3.66 \\
A1835 & rp800569n00 & 6.0 & 0.2523 & 14 01 00.9 & +02 53 10 & 2.3 &
$9.8\pm1.4$ [1] & 30-148.2 & 1.41 & 3.16 \\
A2029 & rp800249n00 & 9.7 & 0.0765 & 15 10 54.7 & +05 45 07 & 3.0 &
$8.5\pm0.2$ [1] & 60-117.3 & 2.17 & 5.24 \\
A2142 & [see note] & 18.9 & 0.0899 & 15 58 18.3 & +27 14 07 & 4.2 &
$9.3\pm0.8$ [1] & 60-134.8 & 2.49 & 3.18 \\
A2163 & wp800385 & 6.8 & 0.2030 & 16 15 44.6 & $-06$ 08 45 &12.1
&$13.8\pm0.5$ [1] & 30-128.0 & 2.37 & 3.77 \\
A2204 & rp800281n00 & 5.2 & 0.1523 & 16 32 46.1 & +05 34 55 & 5.7 &
$9.2\pm1.5$ [1] & 30-103.6 & 1.50 & 9.37 \\
A2218 & rp800097n00 &34.9 & 0.1750 & 16 35 49.3 & +66 12 58 & 3.2 &
$7.1\pm0.2$ [1] & 30-115.0 & 2.01 & 3.68 \\
A2219 & rp800571n00 & 8.0 & 0.2280 & 16 40 17.5 & +46 42 58 & 1.8 &
$12.4\pm0.5$ [1]& 30-138.6 & 2.29 & 2.70 \\
A2244 & rp800265n00 & 2.8 & 0.0970 & 17 02 40.2 & +34 03 37 & 2.1 &
$7.1\pm2.4$ [3] & 30-71.9 & 1.40 & 3.00 \\
A2256 & rp100110n00 &16.1 & 0.0581 & 17 03 08.1 & +78 39 19 & 4.1 &
$7.1\pm0.2$ [5] & 30-45.9 & 2.50 & 2.71 \\
A2319 & rp800073a01 & 2.6 & 0.0559 & 19 21 09.9 & +43 56 58 & 7.9 &
$9.3\pm0.2$ [1] & 60-88.6 & 2.61 & 3.26 \\
A2390 & wp800570n00 & 8.3 & 0.2279 & 21 53 35.1 & +17 42 06 & 6.8 &
$11.1\pm1.0$ [2] & 30-138.6 & 2.01 & 2.92 \\
A2507 & rp800088n00 & 4.8 & 0.1960 & 22 56 49.8 & +05 30 28 & 5.6 &
$9.4\pm1.6$ [3] & 30-124.8 & 1.69 & 2.28 \\
A2744 & rp800343n00 &13.3 & 0.3080 & 00 14 17.5 & $-30$ 23 32 & 1.6
&$11.0\pm0.5$ [1]& 30-167.5 & 1.42 & 3.27 \\
A3112 & rp800302n00 & 6.7 & 0.0746 & 03 17 55.6 & $-44$ 13 56 & 2.6
&$4.1\pm1.4$ [3] & 30-57.4 & 1.35 & 2.99 \\
A3266 & wp800552n00 &12.7 & 0.0594 & 04 31 15.9 & $-61$ 26 48 & 1.6
&$8.0\pm0.3$ [5] & 30-46.8 & 2.04 & 3.53 \\
A3888 & rp700448n00 & 4.0 & 0.1680 & 22 34 26.3 & $-37$ 43 50 & 1.2 &
($9.0\pm1.2$) [6] & 30-111.6 & 1.28 & 5.85 \\
IRAS 09104 & rp701555n00 & 5.8 & 0.4420 & 09 13 45.2 & +40 56 31 & 1.0 &
$8.5\pm3.4$ [1] & 60-404.6 & 1.42 & 2.04 \\
MS 1358 & rp800109n00 & 18.4 & 0.3290 & 13 59 49.4 & +62 31 19 & 1.9 &
$7.5\pm4.3$ [1] & 60-347.9 & 1.22 & 3.44 \\
MS 2137 & rp800573n00 & 8.8 & 0.3130 & 21 40 13.9 & $-23$ 39 29 & 3.6 &
$5.2\pm1.1$ [1] & 30-169.1 & 1.10 & 2.98 \\
PKS 0745 & wp800623n00 & 7.4 & 0.1028 & 07 47 30.1 & $-19$ 17 09 & 46.6
&$8.7\pm1.0$ [1] & 30-75.5 & 1.93 & 2.14 \\
Triang. Aus. & rp800280n00 & 6.4 & 0.0510 & 16 38 20.4 & $-64$ 21 14 &
13.0& $10.1\pm0.7$ [4] & 30-40.8 & 2.41 & 4.27 \\
Zw 3146 & rp800520n00 & 7.9 & 0.2906 & 10 23 39.3 & +04 11 31 & 3.0 &
$11.3\pm3.5$ [1] & 30-161.8 & 1.05 & 2.88 \\
\hline
\end{tabular}
\end{table*}


We have compared samples of clusters of known X-ray luminosity
(David et al. 1993, White et al. 1997, Markevitch 1998, Allen \&
Fabian 1998) with the {\it ROSAT} archive in order to select observations
of bright luminous clusters ($L_X > 10^{45}$ erg s$^{-1}$) at moderate
redshifts ($z>0.05$). This optimises analysis of the distribution
of the gas in the outer regions of each cluster. 

Furthermore, we have not considered clusters that, although matching the
selection criteria presented above, either have evidence of a major merger
that affects both the determination of the centre and the hydrostatic
equilibrium, like Cygnus-A (Owen et al. 1997), A754 (Henriksen \&
Markevitch 1996), A2255 (Davis \& White 1998), A3667 (Rottgering et al.
1997) or are part of larger and more complex system (e.g. A85, Durret et
al. 1998).

Here we note that this sample is not complete in any sense: for
example, if we consider the X-ray-brightest Abell-type clusters sample
(XBACs, Ebeling et al. 1996), that is complete at the 80 per
cent up to redshift of 0.2 for flux in the {\it ROSAT} band (0.1 -- 2.4
keV) larger
than $5 \times 10^{-12}$ erg cm$^{-2}$ s$^{-1}$, and make the same
selection done here, we find 110 items. Of these, 28 are in common
with our sample, 53 are not available in the {\it ROSAT} archive, 19
have only HRI images, 8 have PSPC data but no information on
the gas temperature and 2 are now confirmed with $L_{\rm X, bol} <
10^{45}$ erg s$^{-1}$.

In Table~1, we present the list of the selected clusters with their
basic physical parameters. The intracluster temperatures, $T_{\rm gas}$,
come from published spectral analyses. The clusters in our
sample are very luminous hot objects for which only
observatories with a wide X-ray energy band, like {\it GINGA} and {\it
ASCA}, can properly measure temperatures using the hard tail of the X-ray
spectrum. We therefore use {\it ASCA}
measurements for all the clusters apart from A483 and A2507 ({\it GINGA}
data),  A3112 ({\it EXOSAT}), A545, A2244 ({\it Einstein} MPC), and
A3888 (from its optical velocity dispersion).

In particular, we are interested in the temperature of the bulk of the 
cluster gas, possibly not affected from the presence of any cooling central 
gas, which can lead to a lower (emission-weighted) temperature.
Thus, in the following analysis, we consider gas temperatures that have
measured either excluding the core region (Markevitch 1998)
or including a cooling flow component in the spectral fit (Allen \& 
Fabian 1998).

The error bars on the temperatures quoted in Table~1 are {\it symmetric}
$1 \sigma$ uncertainties. When the published source reports asymmetric
errors at the 90 per cent confidence level, we consider the largest
value and divide by 1.64 (this assumes that the errors are Gaussian).

The {\it PSPC} images have been constructed from counts in the 0.5-2 keV
band, where the Galactic and particle background are minimum, after
correction for instrumental and telemetry dead time, exclusion of times of
high background counts. We also require the Master Veto count rate to be
less than 170 counts s$^{-1}$ (cf. guidelines for reduction of PSPC data
in Snowden et al. 1994). Using the {\it ROSAT} Interactive Data Language
(IDL)  user-supplied libraries, we have divided the images by the
respective exposure maps in the same energy band and corrected them for
vignetting and exposure-time. The region of the detector support rib has
been masked out as well all detected point sources. 
The clusters examined do not show evidence for a major merger, but may be
affected by low level substructure. These clumps have also been masked
despite their small contribution to the total flux. 

The surface brightness profiles have been extracted by estimating the
X-ray centre from the exposure-corrected image after smoothing by a median
filter of width of 5 pixels.
 
The background has been calculated as an average of the counts s$^{-1}$
arcmin$^{-2}$ present between 40$'$ and 45$'$ in a sector which has no
significant contamination from non-cluster emission (cf. column ``bkg" in
Table~1). 

The outer radius, $R_{\rm out}$ (Table~1), of the extracted and
background-subtracted profile is defined as the maximum radius where the
signal exceeds twice the error
present in that radial bin. This error is calculated by adding in
quadrature the Poisson errors on the photon counts both in that bin and
the background.

The bin size, of at least 30 arcsec to avoid any effect of the
Point-Spread-Function, is chosen to improve the signal-to-noise ratio in
the outskirts of the cluster. In clusters where an improvement is
significant, we quote in Table~1 values for bin sizes larger than 30
arcsec.
 
Fig.~\ref{fgas1} plots all the surface brightness profiles (in
counts s$^{-1}$ arcmin$^{-2}$) against radius (in Mpc).
We convert from the angular size to the physical dimension in each cluster,
using the following equation for angular distance
\begin{equation}
r {\rm (Mpc)} =  87.21 \ r {\rm (arcmin)}
\frac{ q_0 z + (q_0 -1) (\sqrt{2q_0 z +1} -1) }{H_0 q_0^2 (1+z)^2}.
\label{fgas:eq3}
\end{equation}
We use, as cosmological parameters ($H_0$, $q_0$), the values (50 km s$^{-1}$
Mpc$^{-1}$, 0.5).

Here we note that this proper radius is inversely proportional to
$H_0$ and depends slightly upon $q_0$ (a variation of 5 per cent is
observed on changing $q_0$ from 0.5 to 0.01 in the redshift range
0.01--0.4, 
with the larger deviation of about 10 per cent at the highest redshifts).
We discuss further in Section~4 the cosmological dependence of the proper
radius, $r$.


\section{DEPROJECTION AND FITTING ANALYSIS}


We analyse the surface brightness profiles to determine the distribution of 
the gas density in the clusters through the two usual techniques, i.e. by
fitting the surface brightness profile and by deprojecting it.
Both techniques make the reasonable assumption that the observed projected
cluster emission is due to X-ray emitting gas which is spherically
symmetric.
In particular, the former assumes a model for the gas density, projects it
on the sky and fits it to the data to constrain the parameters of the
model; the latter, the deprojection technique, makes a proper geometrical
deprojection of the profile and determines the gas density and
temperature profile, assuming a dark matter distribution.

In the next two subsections, we present these two methods and the results
obtained on the distribution of the intracluster gas.
In the last subsection, the constraints on the dark matter in the
case of the hydrostatic equilibrium are discussed.

\subsection{Fitting approach}

We have adopted here the following two models for the gas density
and the projected surface brightness profile:

\begin{enumerate}
\item a single $\beta$-model (Cavaliere \& Fusco-Femiano 1976), 
\begin{equation}
\rho_{\rm gas} = \rho_{0}\left[ 1 + \left(\frac{r}{r_{\rm c}} \right)^2
\right]^{-1.5 \beta} \rightarrow \ S_{\rm b} = S_0 \left[1 +
\left(\frac{r}{r_{\rm c}} \right)^2 \right]^{0.5 -3 \beta};
\label{fgas:eq4}
\end{equation}


\item a gas density profile obtained using the Navarro, Frenk \& White (1997;
hereafter NFW) dark matter profile in the hydrostatic equation where the gas is
assumed isothermal (cf. Appendix; a first application on the Perseus cluster is
presented in Ettori, Fabian \& White 1998; a first theoretical discussion
is found in Makino et al. 1998):
\begin{equation}
\rho_{\rm gas} = a_0 (1+x)^{\eta/x},
\label{fgas:eq5}
\end{equation}
where $x= r/r_{\rm s}$, $\eta = 4\pi G \rho_{\rm s} r^2_{\rm s} \mu
m_{\rm p} / (kT_{\rm gas})$ and $\rho_{\rm s} = \rho_{\rm c} \delta_{\rm
c} (1+z)^3 \Omega_0/\Omega_{\rm z} $, with $\delta_{\rm c}$ equal to the
characteristic density of the cluster and $\rho_{\rm c}$ to the critical
density (see Appendix in Navarro, Frenk \& White 1997). 

The surface brightness profile is then obtained by numerical
integration of the gas density through equation~\ref{fgas:eq1}.

\end{enumerate}

All the fits have been performed by a non-linear least squares technique
(routine {\it Curvefit} in IDL). This algorithm, however, is 
sensitive to large departures for a (generally small) number
of data points, the so-called {\it outlier points}. A statistical
estimator that is able to properly weight these points is qualified
as robust.
Thus, we have also tested our results with a robust estimator of
the minimization of the function $(y [{\rm raw \ data}] - y [{\rm
fit}])/\sigma[{\rm raw \ data}] $, namely the downhill simplex method
implemented in the {\it amoeba} routine (Press et a. 1992; IDL vers.~5.0).
Generally, the agreement is good (with a deviation of the best estimate of
the parameters of about 1 per cent in average) due to the large number of
radial bins available. 

\begin{figure*}
\hbox{
\psfig{figure= 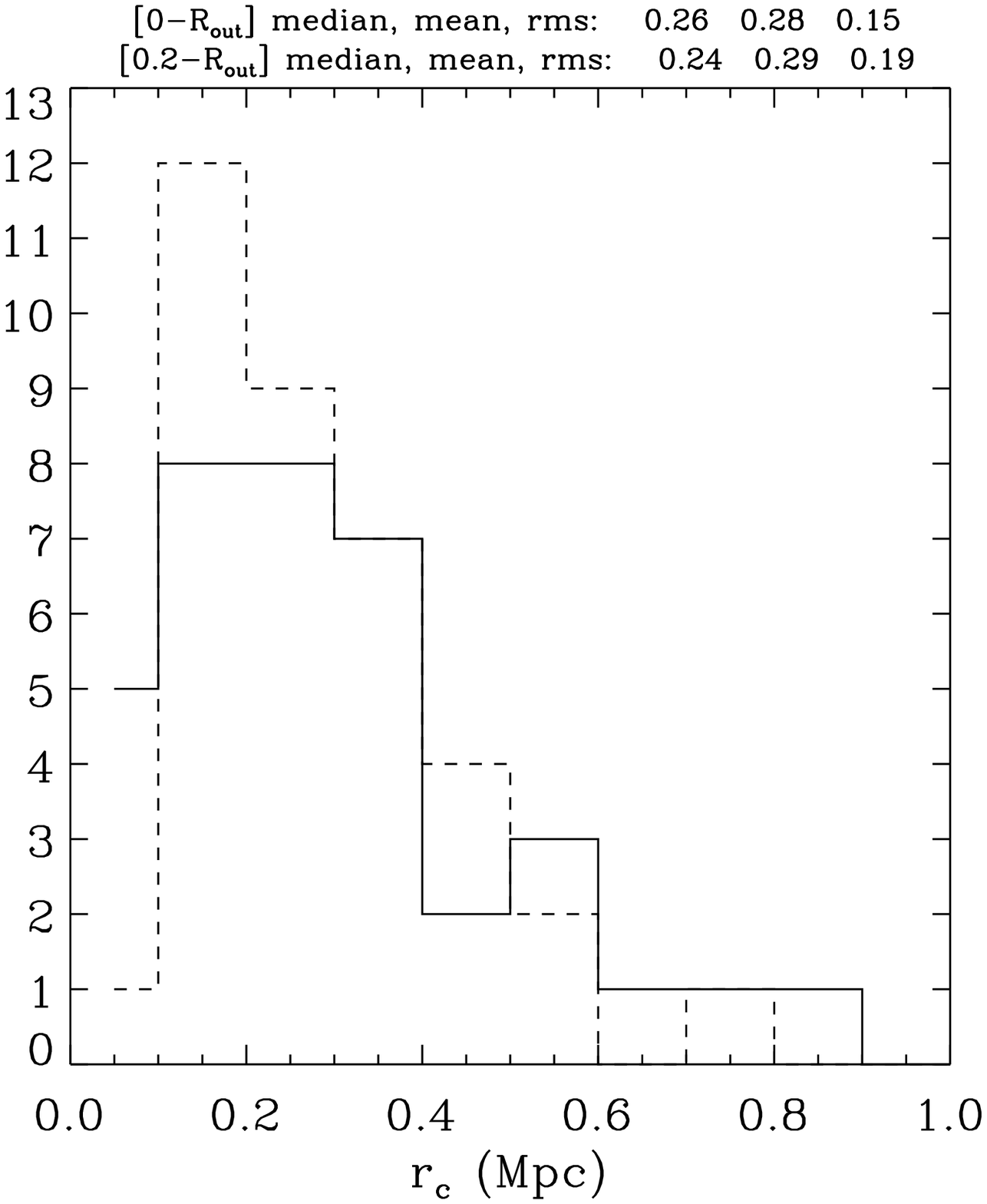,width=.5\textwidth,angle=0}
\psfig{figure= 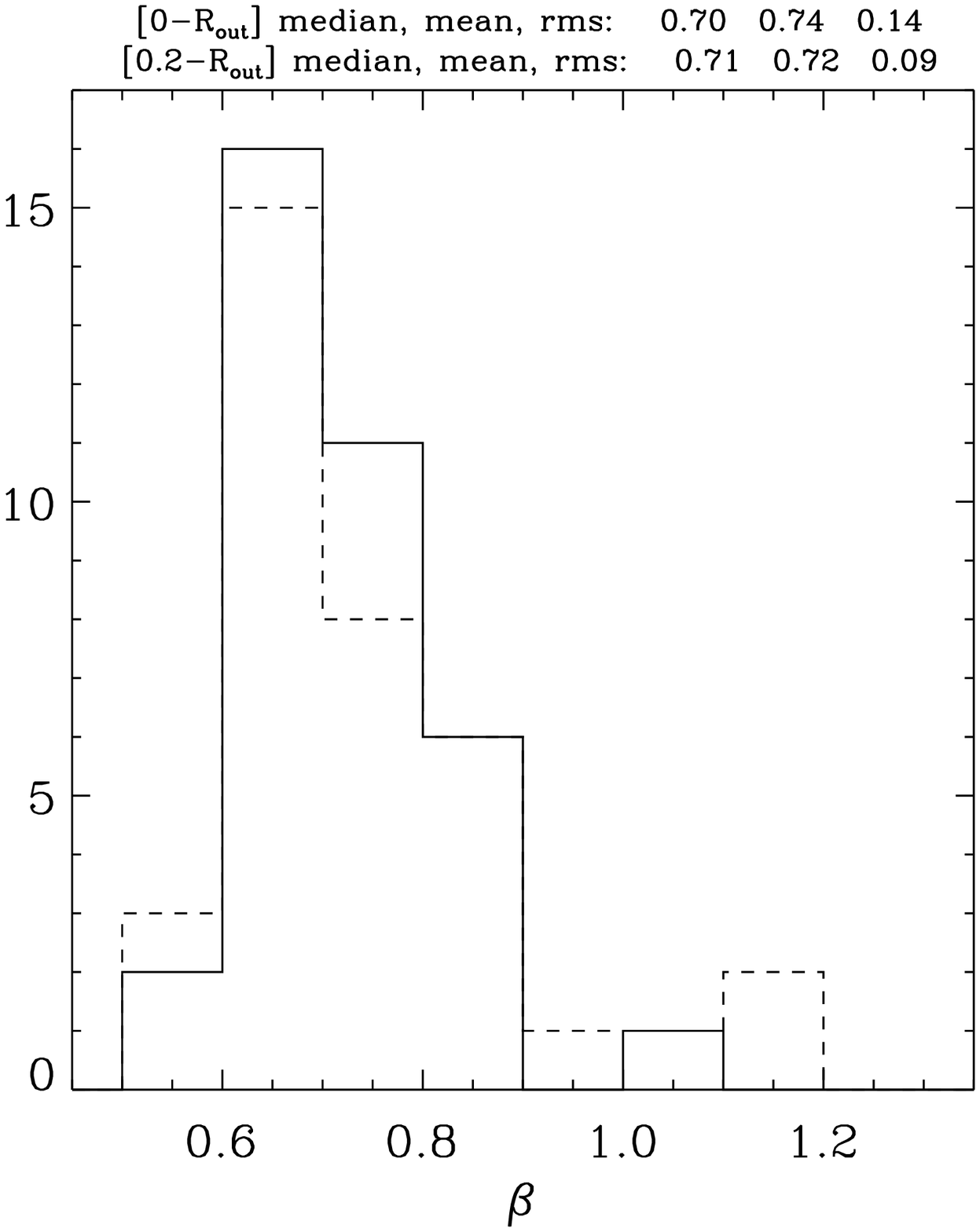,width=.5\textwidth,angle=0} }
\hbox{
\psfig{figure= 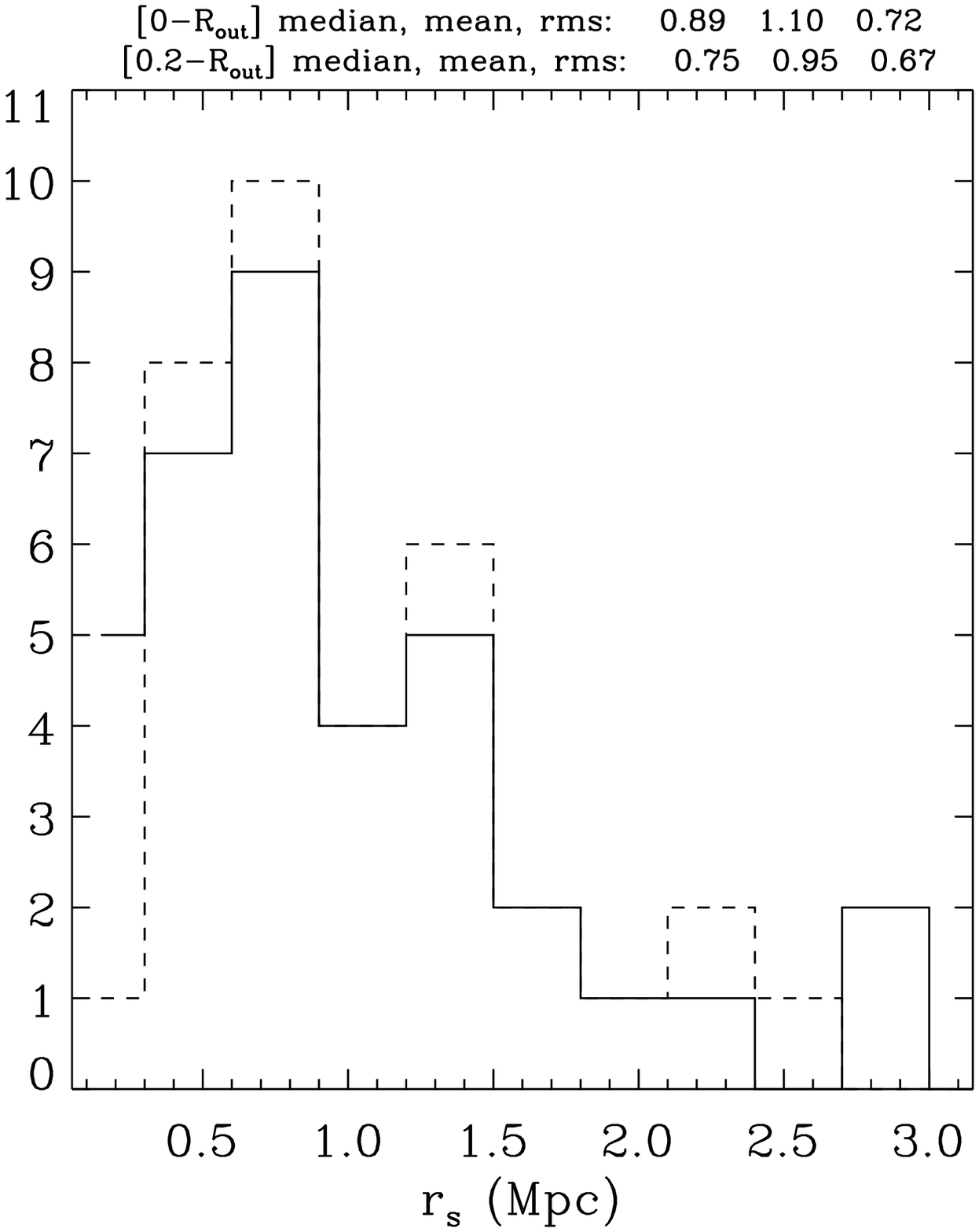,width=.5\textwidth,angle=0}
\psfig{figure= 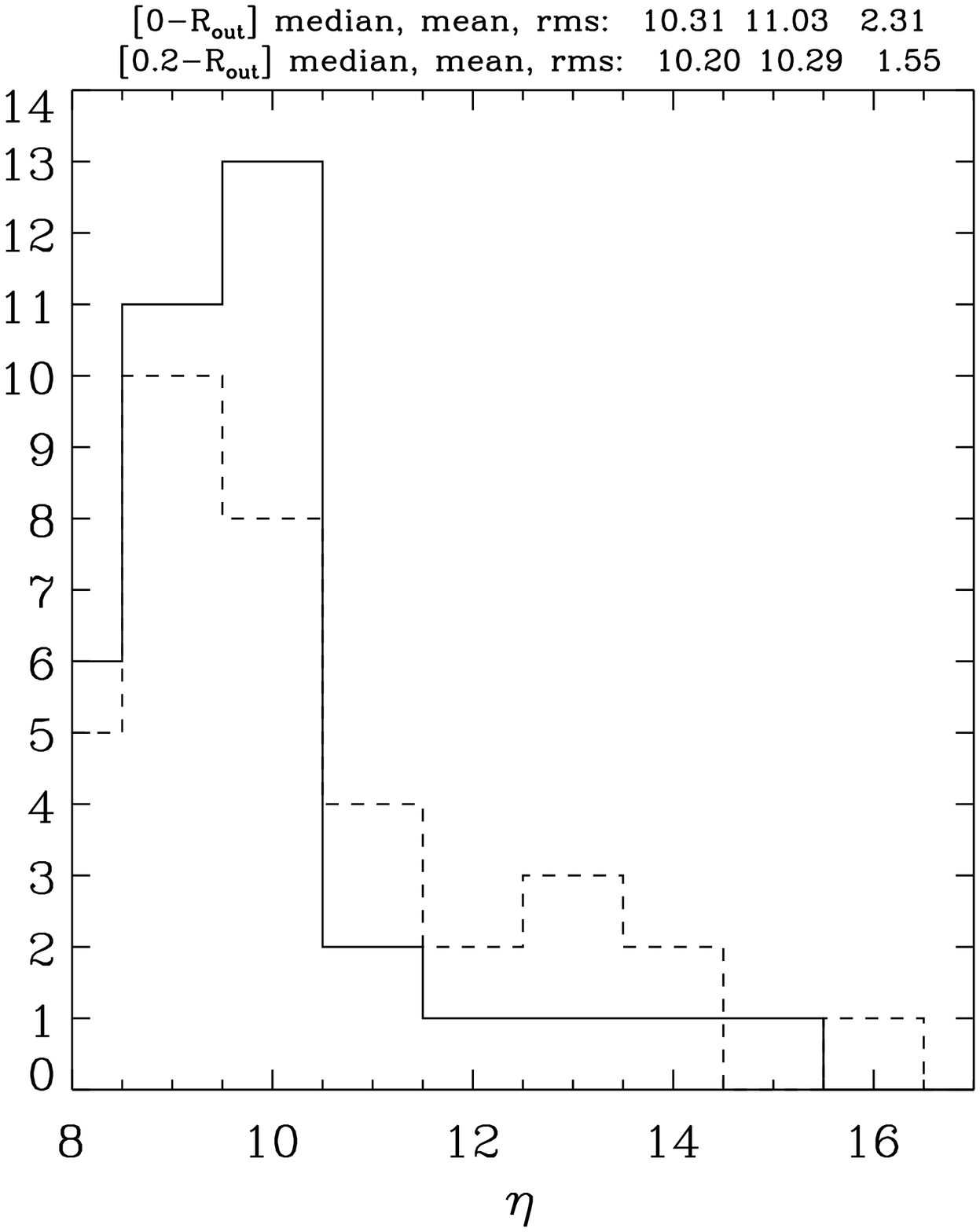,width=.5\textwidth,angle=0}
} \caption{ Histograms of the parameters from the $\beta$-model and 
NFW gas profile. The histograms of the best-fit parameters in the range
$[0.2 - R_{\rm out}]$ and $[0 - R_{\rm out}]$ are with {\it solid} and
{\it dashed} line, respectively.
The median, mean and standard deviation for each 
distribution is quoted in the title of each plot. 
} \label{fgas2} \end{figure*}

In the following analysis, we present the results obtained from the robust
estimate of the model parameters.
We show in Fig.~\ref{fgas2} the count distribution of the values of the
$\beta$-model parameters, $(r_{\rm c}, \beta)$, and for the NFW gas 
profile, $(r_{\rm s}, \eta)$. In this figure, we overplot the best-fit
results obtained over the radial range $[0, R_{\rm out}]$ and 
$[0.2, R_{\rm out}]$ Mpc. The choice of the second radial range avoids
any cooling flow (if present in the inner part of the cluster) on the
estimates of the parameters. Doing this, we measure [average, dispersion]
best-fit values of $r_{\rm c}$ = [0.29, 0.19] Mpc, $\beta$ = [0.72, 0.09],
and $r_{\rm s}$ = [0.95, 0.67] Mpc, $\eta$ = [10.29, 1.55]. 
 
A range of tests to check whether the two sample
populations, histograms for which are plotted in Fig.~\ref{fgas2}, have
significantly different mean and/or variance, shows disagreement
at 95 per cent confidence level of the variance (``F-variance
test'') in the populations of $\beta$ and $\eta$.

\begin{figure*}
\hbox{ \psfig{figure= 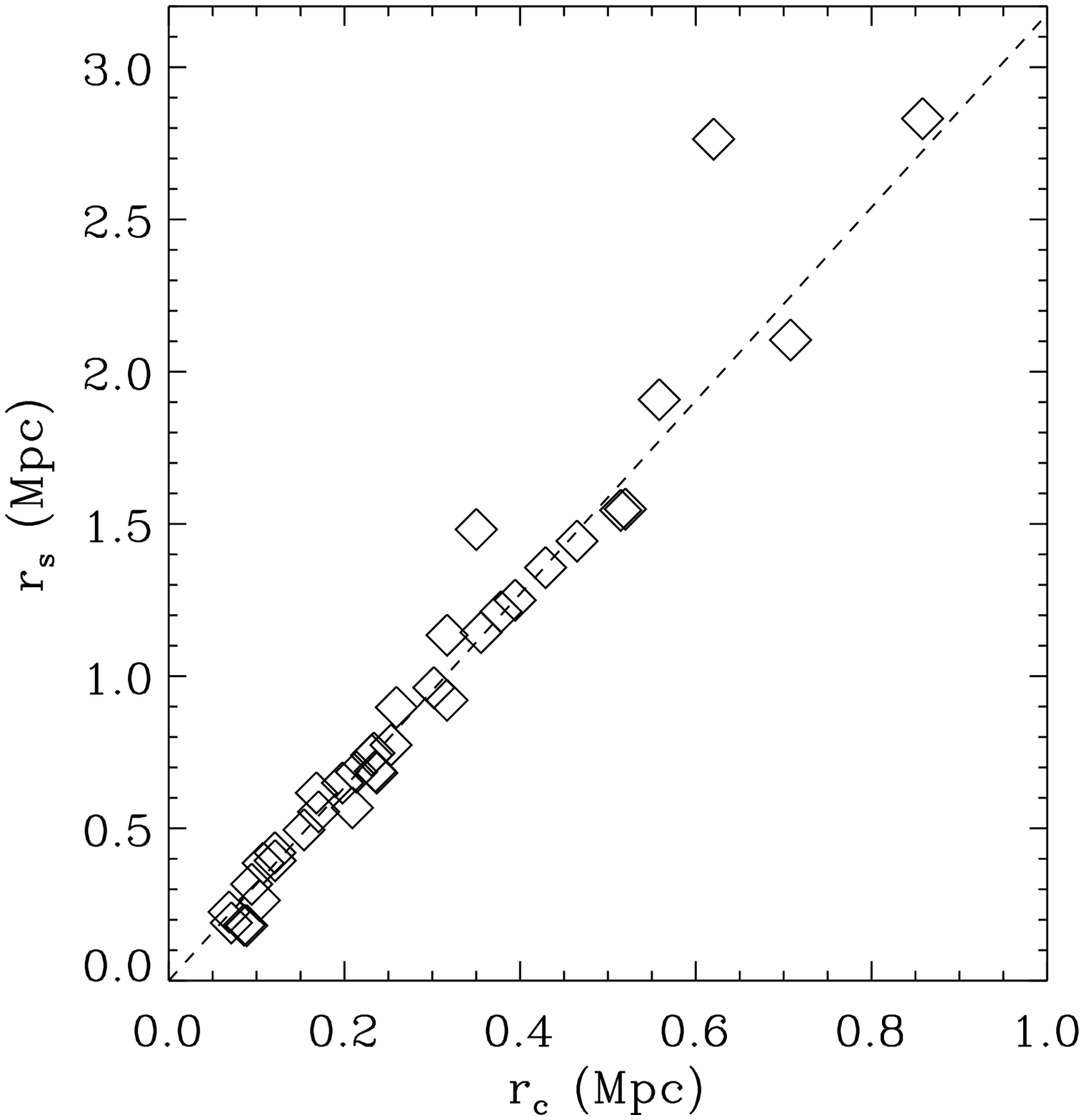,width=.5\textwidth,angle=0}
\psfig{figure= 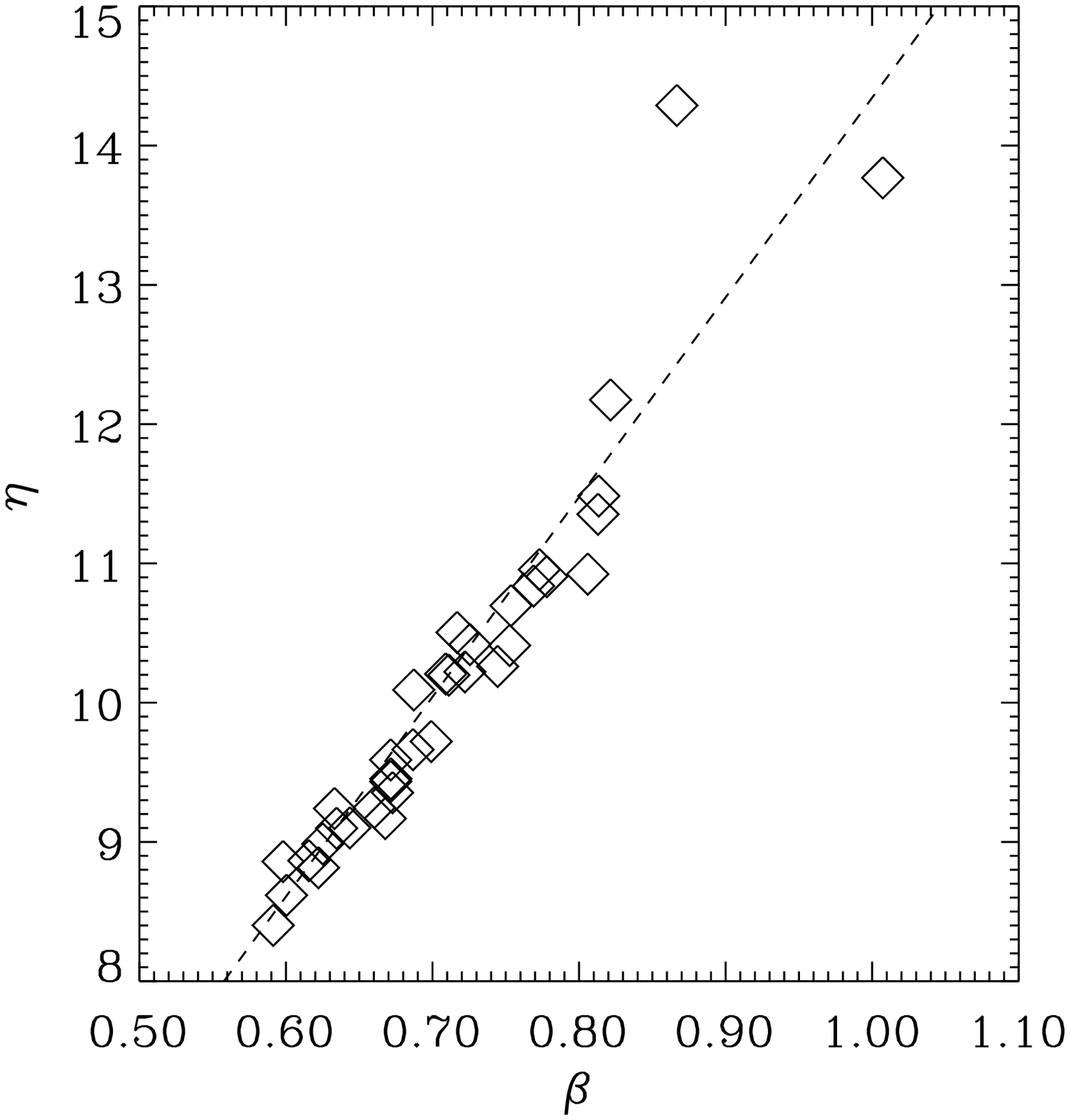,width=.5\textwidth,angle=0} }
\caption{Correlations between the scale and slope parameters in the 
$\beta$-model and NFW gas profile.
} \label{fgas3} \end{figure*}

With the intention to link the results of the fit using both the
$\beta$-model and the NFW profile, we perform a linear
unweighted polynomial fit on the grid of parameters (Fig.~\ref{fgas3}). 
We obtain $r_{\rm s} = 3.17 \ r_{\rm c}$ and $\eta = 14.34 \ \beta$.
These correlations are consistent with the best-fit results, when one
model is fitted with the other (cf. also Makino et al. 1998).

In the following analysis, we use the results from the fit with the NFW
gas profile. A comparison between the two fits shows that the NFW
gas profile provides a $\chi^2$ lower than the value from the
$\beta-$model in 19 out of 36 clusters. 
Furthermore, even if the fit with the NFW gas profile is computationally
more expensive, we show in the Appendix that its best-fit parameters are
directly linked with the scale radius and normalization of the dark matter
profile obtained from N-body simulations. 

In particular, as we discuss in Section~3.3, we are interested in defining
in each cluster the radius, $r_{500}$, where the overdensity of the dark
matter with respect to the average value is 500. Hence, after we have
measured the best-fit parameters in $[0.2, R_{\rm out}]$ Mpc, we calculate
$r_{500}$ and, for only those clusters in which $r_{500} < R_{\rm out}$,
reiterate the fit over the range $[0.1, 1] \ r_{500}$ until the
convergency on $r_{500}$ is reached, i.e. the estimated value of $r_{500}$
is within 0.01 Mpc. A maximum of 3 trials is required to converge. 

The corresponding best fit values are quoted in Table~2 and then used 
to describe the gas distribution through the deprojection of the central
density. Using XSPEC (vers.10, Arnaud 1996), we
have converted the fitted count rates to the flux due to thermal
X-ray emission from an optically-thin plasma [{\it MEKAL} code,
based on the model calculations of Mewe
and Kaastra (Kaastra 1992) with Fe L calculations by Liedahl (1995)],
assumed isothermal at the temperature given in Table~1, with a fixed
metallicity of 0.4 times solar abundance, and with Galactic absorption
included (cf. Table~1).

\begin{figure}
\psfig{figure= 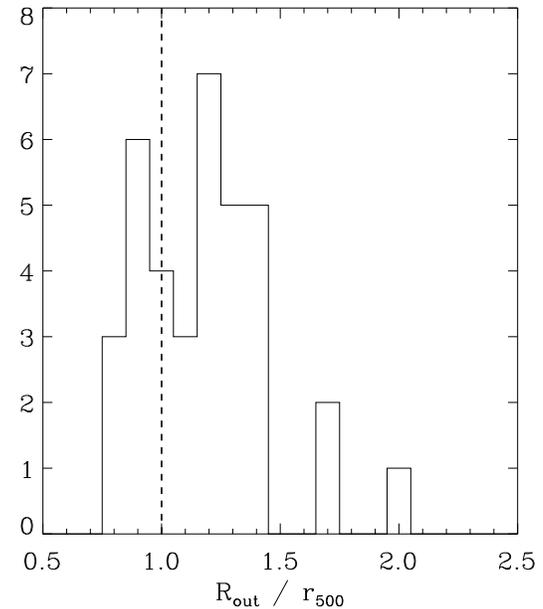,width=.5\textwidth,angle=0}
\caption{The histogram for $R_{\rm out}/r_{500}$ provides an indication of 
the region `seen' with respect to the extrapolated one. Values for this
ratio greater than, or equal to, 1 confirm that our profiles can trace
the mass within an overdensity in dark matter of 500. This happens in 23
of the 36 clusters (64 per cent). } \label{fgas4}  \end{figure}

\subsection{Deprojection procedure}

In the deprojection technique, the count emissivity in each radial
volume shell is calculated analytically and compared with the
predicted counts from the emission predicted from an optically thin gas
(described by a {\it MEKAL} model) absorbed by intervening matter ($N_{\rm
H}$ from Table~1) and convolved with the response of the detector. 
After selecting boundary condition
(i.e. the pressure in the outermost bin which allows the resulting
deprojected temperature profile to match the observationally determined
cluster temperature), the gas temperature and density profiles are
obtained,  once a model for the dark matter distribution is defined 
(see White, Jones \& Forman 1997 for a detailed discussion
on the deprojection technique).

The gravitational potential is described by a functional form given by the
sum of two contributions: the central galaxy potential parametrised by a
de Vaucoulers law (1948), assumed fixed for all the clusters and
with parameters $R_{\rm eff} = 30$ kpc and velocity dispersion of 300 km
s$^{-1}$ (Malumuth \& Kirshner 1985), and a Navarro-Frenk-White potential
for the general cluster. The latter potential has a scale parameter,
{\it core}, and a velocity dispersion, $\sigma_{\rm DM}$, which is well
represented, under the isothermal assumption, by the intracluster temperature
[$\sigma_{\rm DM} = (k T_{\rm gas}/ \mu m_{\rm p})^{0.5}$; cf. Table~2].

For some clusters with $T_{\rm gas}$ larger than 10 keV, for which the
temperature determination is not very precise (e.g.  A2163, A2744), we fix
$\sigma_{\rm DM}$ to the optical estimate of the velocity dispersion and
require a flat temperature profile.  The required values for the velocity
dispersion are larger than the predicted one from the isothermal relation
by 15 and 50 per cent, for A2163 and A2744, respectively. 

\subsection{The total gravitating mass}

\begin{figure}
\psfig{figure=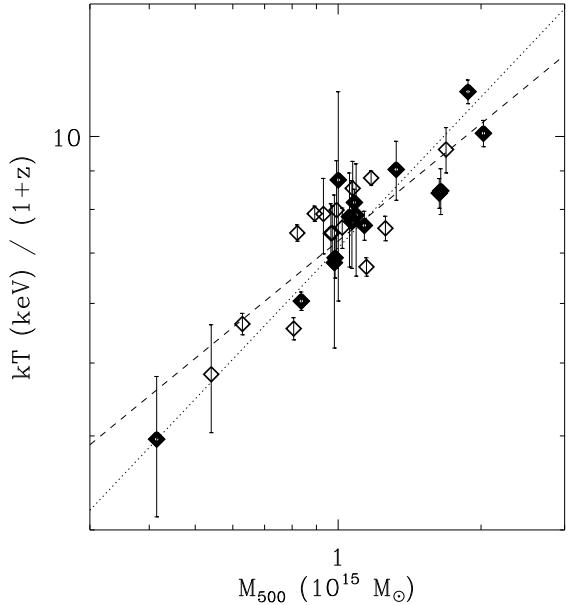,width=0.5\textwidth,angle=0}
\caption{This plot shows the gas temperature vs. $M_{500}$ with the best-fit
power law functions: ({\it dotted line}) $M_{500} = 1.67 (\pm 0.02)$
(T/10 keV)$^{1.5}$ $10^{15} M_{\odot}$ ($\chi^2 = 212$, with 29
degree-of-freedom); ({\it dashed line}) $M_{500} = 1.86 (\pm 0.04)$ (T/10
keV)$^{1.93\pm0.09} 10^{15} M_{\odot}$ ($\chi^2 = 184$, with 28
degree-of-freedom). The clusters at high redshift, with respect to the
median value, are the solid symbols. 
} \label{fgas5} \end{figure}

From the fits on the surface brightness profile, and the deprojections, 
we obtain constraints on the gas density distribution under the
assumption of an isothermal plasma in a spherically symmetric cluster. 

To do this, we have assumed a functional form for the total matter
density and applied the hydrostatic equilibrium written in
equation~\ref{fgas:eq2}.
The distribution of the total gravitating mass is then obtained through
the radial integration of the total matter density profile.


We recall that our estimates of the total mass, $M_{\rm tot} (<r)$, are
based upon (i) radial surface brightness profiles that reach a median
$R_{\rm out}$ of 1.93 Mpc, (ii) gas temperatures that both are corrected
for any cool component in the core and were measured from
broad-band detectors that describe well the shape of the thermal
emission from the clusters, (iii) the assumption that the gas is
isothermal (but see considerations in Sect.~4.3).

This last assumption seems reasonable and conservative from recent results
on simulations of gas dynamics (Evrard, Metzler \& Navarro 1996), 
if one considers regions
of the clusters where the overdensity of dark matter with respect to the
average value is 500.  Thus, we quote in the following analysis the value
of the radius, $r_{500}$, where this overdensity is reached, and $M_{\rm
tot} (< r_{500}) = M_{500}$ using the best-fit parameters of the NFW gas
density profile and the relations discussed in the Appendix. 
 
Considering that we obtain a median value for $r_{500}$ of 1.66 Mpc to be
compared to a median $R_{\rm out}$ of 1.93 Mpc (cf. Fig.~\ref{fgas4} ), 
the proper gravitating mass for each cluster is here
determined in a robust way, generally without any extrapolation or 
application of either the $r-T_{\rm gas}$ or $M_{\rm tot}-T{\rm gas}$
relation (Evrard
et al. 1996, Hjorth, Oukbir \& van Kampen 1998), which would need a
proper calibration through independent measurements of the mass  (e.g.
hydrodynamics simulations, gravitational lensing) and replaces
the peculiarity of a cluster with an average behaviour.


For comparison with previous work, we have also investigated the $M_{500}
- T_{\rm gas}$ relation, after calculating $M_{500}$ for the 30 clusters
of our final sample (see next section). Adopting the scaling relation $M 
\propto r \ T_{\rm gas} \propto T_{\rm gas}^{3/2} (1+z)^{-3/2}$, 
where the last step implies that $r \propto M^{1/3} (1+z)^{-1}$, 
we obtain a best-fit of $M_{500} = 1.67
(\pm 0.02)  [T_{\rm 10, gas}/ (1+z) ]^{1.5} \times 10^{15} M_{\odot}$,
where $T_{\rm 10, gas}$ is in units of 10 keV (Fig.~\ref{fgas5}). 
Leaving as free parameter the slope of the temperature, we measure
$M_{500} = 1.86 (\pm 0.04) [T_{\rm 10, gas}/ (1+z) ]^{1.93\pm0.09}$,
slightly steeper than a power law with index of 1.5.

At $r_{500}$, Evrard (1997) measures from simulated clusters a coefficient
$2.22 \pm 0.32$ for the above relation. 
The disagreement with our best-fit result is due mainly, apart from
the large scatter, to the steeper gas profiles 
predicted by the simulations.


\section{THE GAS FRACTION}

\begin{figure*}
\psfig{figure=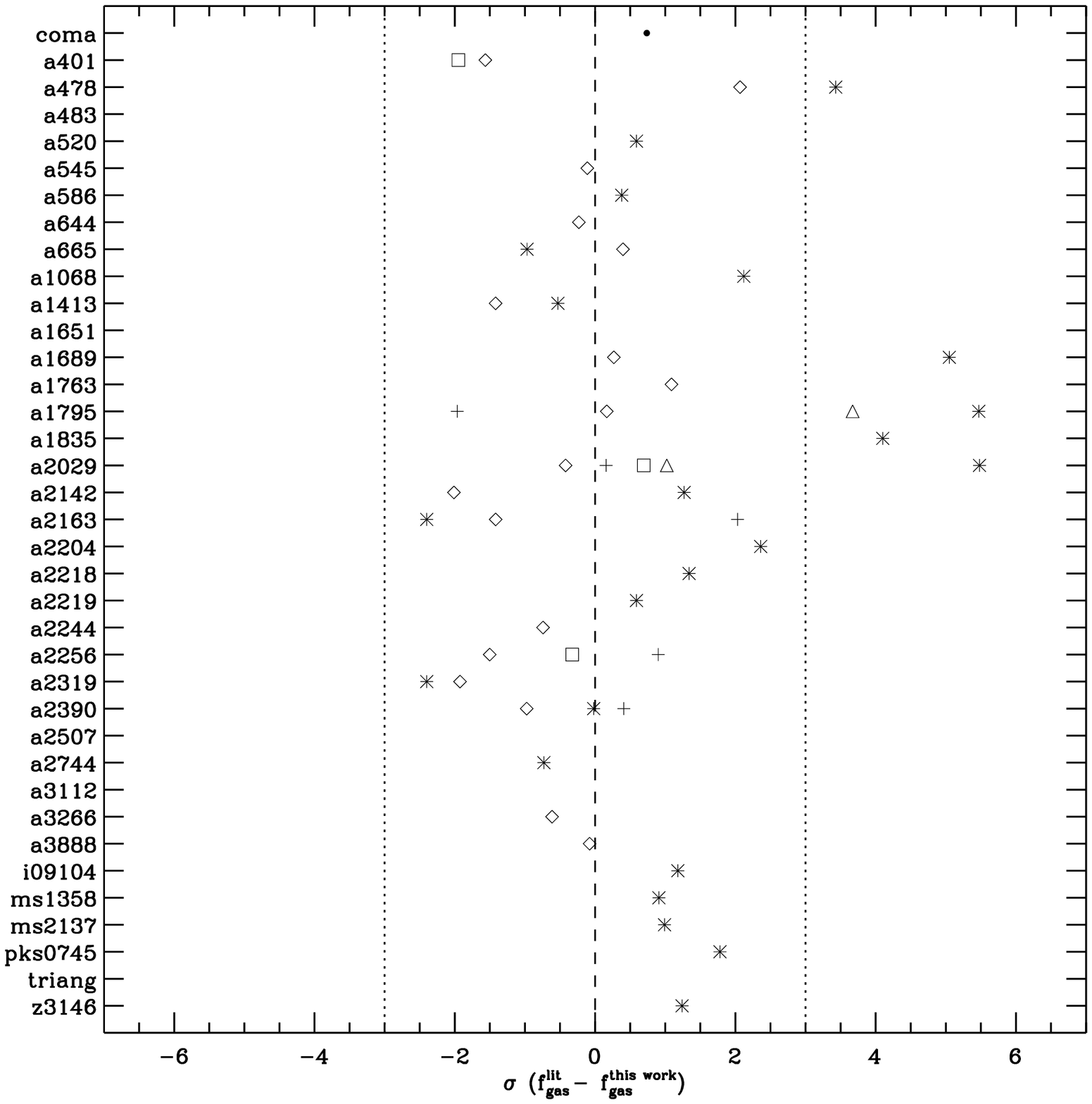,width=\textwidth,angle=0}
\caption{Plot of the difference in standard deviation between
$f_{\rm gas}$ from published sources and our estimate for each cluster in
our sample. 
We compare the data at the same radius and using the same technique 
[i.e. the deprojection for the results from Allen \& Fabian (1998) and
White et al. (1997)]. For the White et al. (1997) results, we propagate
the uncertainties on the ``reprojected" gas temperature to the total mass.
We include also the Coma cluster as comparison.  Note: $\bullet$, White et
al. (1993; at $r=$ 3 Mpc); $+$, A1795:  Briel \& Henry (1996; at $r=$ 1
Mpc), A2029: Sarazin et al. (1998; at $r=$ 1.88 Mpc), A2163: Elbaz,
Arnaud \& B\"ohringer (1995; at $r=$ 1.5 Mpc), A2256: Markevitch \&
Vikhlinin (1997; at $r=$ 3 Mpc), A2390: B\"ohringer et al. (1998; at $r=$
1 Mpc); $\ast$, Allen \& Fabian (1998; at $r=$ 0.5 Mpc); $\Diamond$,
White, Jones \& Forman (1997, up-to-date version of White \& Fabian 1995; 
at $r=$ 1 Mpc); $\triangle$, David, Jones \& Forman (1995; at $r= r_{\rm
500}$); $\Box$, Buote \& Canizares (1996, but after corrections on the
total masses [Buote, priv. comm.]; at $r=$ 1 Mpc). } 
\label{fgas6}
\end{figure*}

\begin{figure}
\psfig{figure= 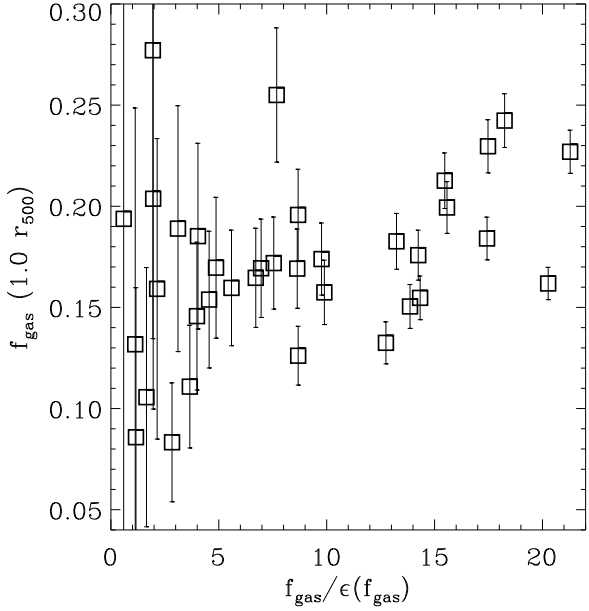,width=.5\textwidth,angle=0}
\caption{The gas mass fraction values for our 36 clusters are plotted
versus the ratio $f_{\rm gas} / \epsilon(f_{\rm gas})$. Only the 30
clusters with $f_{\rm gas} / \epsilon(f_{\rm gas}) > 2$ have been
considered in our final sample.} 
\label{fgas6bis}  \end{figure}

\begin{figure}
\psfig{figure= 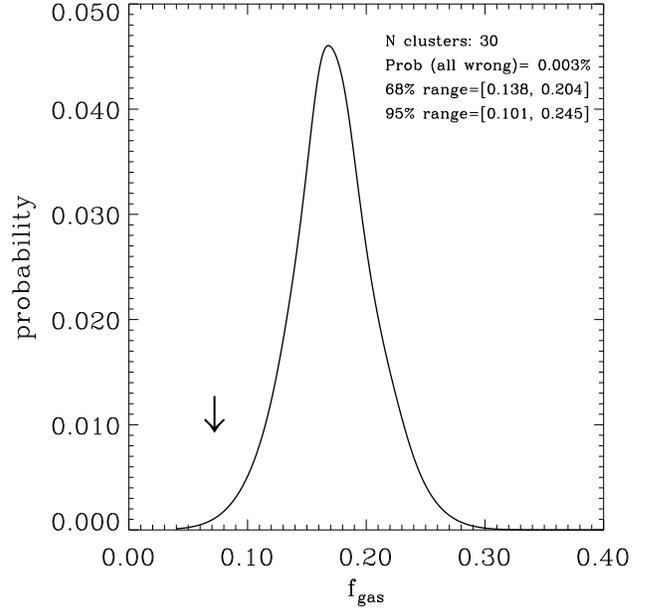,width=0.5\textwidth,angle=0}
\caption{Bayesian probability distribution for the gas fraction observed
in the clusters in our refined sample. This distribution peaks at
$f_{\rm gas} = 0.168$. The arrow indicates the constraints from
the primordial nucleosynthesis value for the low D/H case. It has a
probability of $7.2 \times 10^{-3}$ with respect to the plotted
distribution.
} \label{fgas7} \end{figure}

\begin{figure}
\psfig{figure= 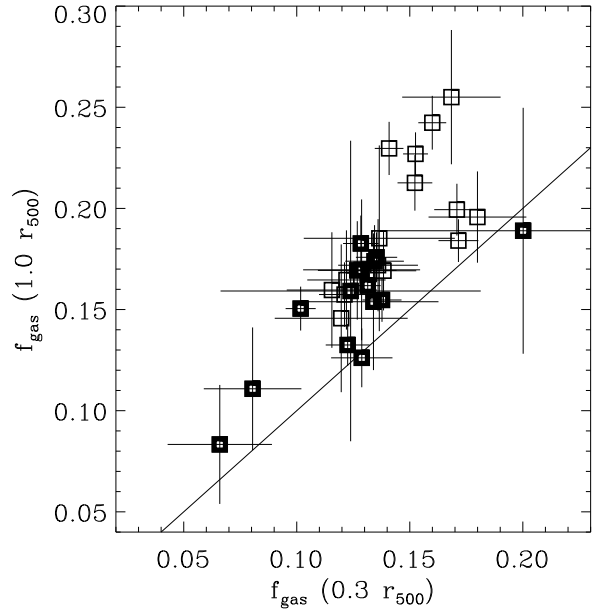,width=0.5\textwidth,angle=0}
\caption{The gas fraction compared at different radii, i.e. at 0.3
and 1 $r_{500}$. This corresponds to about 0.5 and 1.7 Mpc, if we
consider the median value of our sample of 36 clusters. The solid line
indicates $f_{\rm gas} (0.3 \ r_{500}) = f_{\rm gas} (1.0 \ r_{500})$.
Most of the clusters show an excess in the gas contribution at larger
radii. The clusters at high redshift, with respect to the 
median value, are the solid symbols.
} \label{fgas8} \end{figure}

\begin{figure}
\psfig{figure= 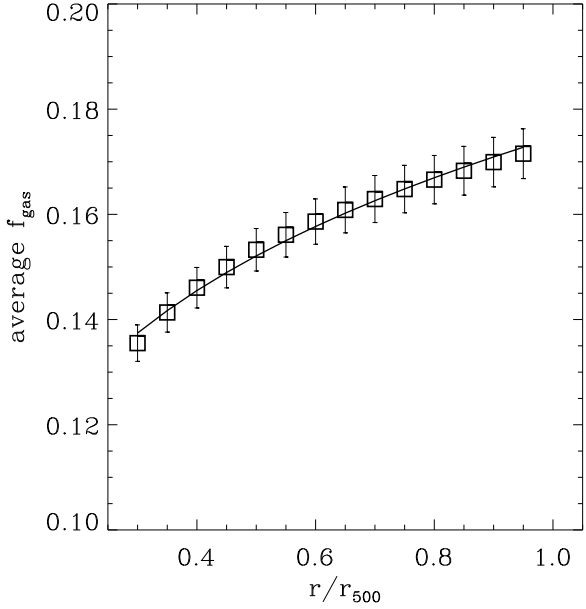,width=0.5\textwidth,angle=0}
\caption{The average at different radii of $f_{\rm gas}$ from all 30
clusters with the respective error. We overplot the best-fit
power law (index of $0.20\pm0.02$; $\chi^2 = 0.77$ for 12
degrees-of-freedom). 
} \label{fgas9} \end{figure}

\begin{figure}
\psfig{figure= 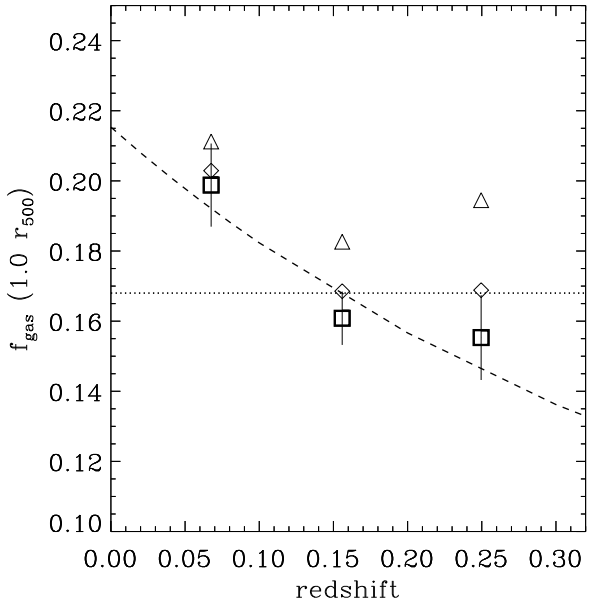,width=0.5\textwidth,angle=0}
\caption{This plot shows the dependence of $f_{\rm gas}$ on the redshift
with the best-fit result of $0.215 (-0.019, +0.020) \ (1+z)^{-1.75 (-0.65,
+0.65)}$ ($\chi^2 = 1.62$, with 1 degrees-of-freedom). 
The $x$-position of the bins is a biweight location.
The $y$-position is the biweighted value with the respective error.
When a different cosmology is considered into the dependence of $f_{\rm gas}$ 
upon $d_{\rm ang}^{1.5}$, we obtain the values represented with {\it diamond}
($\Omega_{\rm 0, m} = 0.2, \Omega_{\Lambda} = 0$) and {\it triangle} 
($\Omega_{\rm 0, m} = 0.2, \Omega_{\Lambda} = 1 - \Omega_{\rm 0, m}$).
The dotted line represents the central value of $f_{\rm gas}$ from the
robust analysis on 30 clusters.
} \label{fgas10} \end{figure}

\begin{figure}
\psfig{figure= 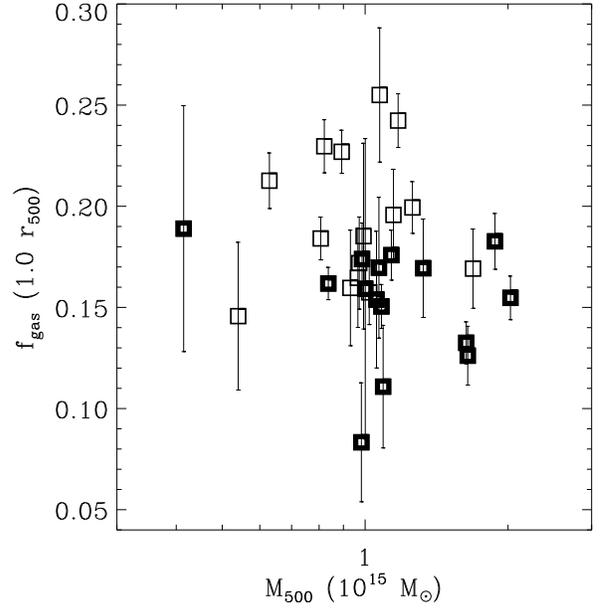,width=0.5\textwidth,angle=0}
\caption{The gas fraction at $r_{500}$ is shown with the corresponding
gravitating mass at the same radius. The clusters at high redshift, with
respect to the median value, are the solid symbols.
} \label{fgas11} \end{figure}

\begin{figure}
\psfig{figure= 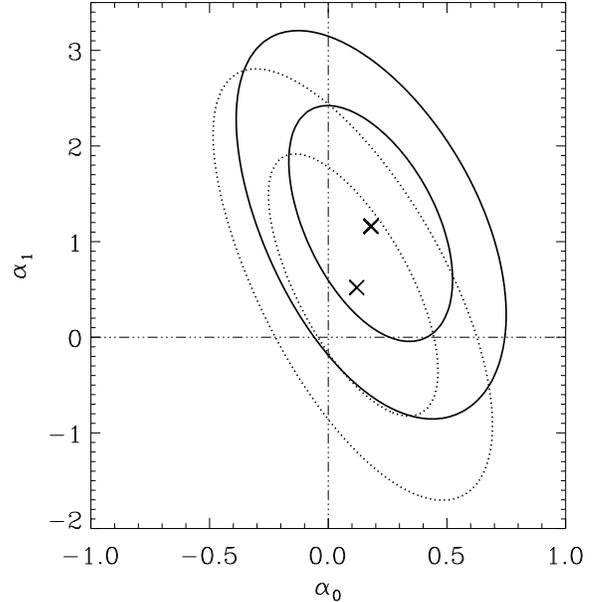,width=0.5\textwidth,angle=0}
\caption{Error contours (68 and 95 per cent confidence, i.e. $\Delta
\chi^2 = 2.30, 6.17$) for the two interesting parameters $\alpha_0$ and
$\alpha_1$ in equation~\ref{fgas:eq6}. {\it Solid line}: $\Omega_{\rm 0,
m} = 1$; {\it dotted line}: $\Omega_{\rm 0, m} = 0.2, \Omega_{\Lambda} = 1
- \Omega_{\rm 0, m}$. 
} \label{fgas12} \end{figure}

\begin{table*}   
\caption{This table contains the dark matter parameters used in the
deprojection analysis, the best-fit parameters from the fitting 
procedure, the value of $r_{500}$ as measured applying the NFW gas
profile, and the gas fraction from the deprojection analysis
calculated, where available (i.e. for radii less than $R_{\rm out}$; cf.
Table~1), at the bin closest to the quoted radius. In {\it italics} we
quote $f_{\rm gas}$ as measured at that radius by the fitting method.
The value of $f_{\rm gas} (r_{500})$ is obtained by using
the best-fit parameters with the NFW profile. 
{\it Note:} {\it core}, $r_{\rm c}, r_{\rm s}, r_{500}$ in Mpc; 
$\sigma_{\rm DM}$ in km s$^{-1}$. In the first column, we quote in
parenthesis $\sigma_{\rm DM}$ as calculated from $T_{\rm gas}$ assuming
the isothermal condition. $^{\dagger}$ from optical analysis in Struble \&
Rood (1991; A2163), Smail et al. (1997; A2744). } 
\begin{tabular}{| l@{\hspace{.7em}} c@{\hspace{.7em}} c@{\hspace{.7em}} 
c@{\hspace{.7em}} c@{\hspace{.7em}} c@{\hspace{.7em}} c@{\hspace{.7em}} 
c@{\hspace{.7em}} c@{\hspace{.7em}} |} \hline
cluster & DEPROJ & \multicolumn{3}{c}{FITTING}  &
\multicolumn{4}{c}{$f_{\rm gas}$} \\
  &  {\it core}, $\sigma_{\rm DM}$  &  $r_{\rm s}, \eta$  & $r_{\rm c},
\beta$ & $r_{500}$  &  $<0.5$ Mpc &  $< 1.0$ Mpc & $< 1.5$ Mpc  
& $< r_{500}$ \\  \hline
\multicolumn{9}{c}{  }\\
A401 & 0.90 1100 ({\it 1103}) & 0.62  8.17 & 0.22 0.59 & 1.66&
0.160 0.007 {\it (0.141)} & 0.194 0.010 {\it (0.187)} & 0.223 0.013 {\it (0.220)} & {\it 0.230 0.013}\\
A478 & 0.40 1100 ({\it 1110}) & 0.46  9.56 & 0.17 0.69 & 1.73& 
0.130 0.017 {\it (0.135)} & 0.157 0.020 {\it (0.153)} & 0.171 0.022 {\it (0.166)} & {\it 0.172 0.023}\\
A483 & 0.80 1150 ({\it 1150}) & 1.13 10.50 & 0.32 0.72 & 1.47&   
0.095 0.033 {\it (0.069)} & 0.106 0.037 {\it (0.081)} & ... &{\it 0.083 0.029}\\
A520 & 1.10 1280 ({\it 1123}) & 1.70 11.35 & 0.50 0.81 & 1.57& 
0.136 0.013 {\it (0.138)} & 0.179 0.017 {\it (0.168)} & 0.196 0.020 {\it (0.174)} & {\it 0.174 0.018}\\
A545 & 0.50  950 ({\it 914}) & 1.51 12.52 & 0.40 0.82 & 1.45&
0.179 0.303 {\it (0.182)} & 0.204 0.345 {\it (0.199)} & 0.243 0.411 {\it (0.193)} & {\it 0.194 0.328}\\
A586 & 0.30 1150 ({\it 1275}) & 0.26  8.81 & 0.10 0.62 & 1.60&
0.072 0.063 {\it (0.083)} & 0.091 0.080 {\it (0.105)} & ...  & {\it 0.132 0.117}\\
A644 & 0.50 1100 ({\it 1110}) & 0.59  9.38 & 0.22 0.69 & 1.79& 
0.103 0.010 {\it (0.119)} & 0.129 0.012 {\it (0.139)} & 0.141 0.014 {\it (0.151)} & {\it 0.157 0.016}\\
A665 & 1.20 1350 ({\it 1170}) & 1.49 10.69 & 0.49 0.74 & 1.67& 
0.133 0.009 {\it (0.135)} & 0.163 0.011 {\it (0.166)} & 0.185 0.013 {\it (0.175)} & {\it 0.176 0.012}\\
A1068 & 0.25  915 ({\it 914}) & 0.42  9.72 & 0.12 0.70 & 1.36& 
0.102 0.025 {\it (0.124)} & 0.119 0.030 {\it (0.138)} & ... &{\it 0.146 0.037} \\
A1413 & 0.45 1140 ({\it 1137}) & 0.57  9.17 & 0.21 0.67 & 1.65&
0.110 0.016 {\it (0.122)} & 0.130 0.019 {\it (0.145)} & 0.143 0.021 {\it (0.161)} & {\it 0.165 0.025}\\
A1651 & 0.40  940 ({\it 963}) & 0.58  9.01 & 0.22 0.66 & 1.50& 
0.127 0.006 {\it (0.158)} & 0.156 0.008 {\it (0.190)} & 0.191 0.012 {\it (0.212)} & {\it 0.213 0.014}\\
A1689 & 0.45 1300 ({\it 1233}) & 0.92 11.35 & 0.32 0.81 & 1.90& 
0.105 0.011 {\it (0.126)} & 0.123 0.013 {\it (0.132)} & 0.134 0.015 {\it
(0.129)} & {\it 0.126 0.015}\\
A1763 & 0.75 1260 ({\it 1214}) & 1.16  9.20 & 0.40 0.66 & 1.64& 
0.090 0.006 {\it (0.102)} & 0.123 0.008 {\it (0.132)} & 0.146 0.010 {\it (0.147)} & {\it 0.150 0.011}\\
A1795 & 0.50  920 ({\it 947}) & 0.77 10.70 & 0.25 0.75 & 1.67&
0.160 0.008 {\it (0.172)} & 0.192 0.010 {\it (0.183)} & ... & {\it 0.184
0.011}\\
A1835 & 0.25 1220 ({\it 1221}) & 0.32 10.22 & 0.09 0.72 & 1.55&
0.112 0.024 {\it (0.135)} & 0.124 0.027 {\it (0.143)} & ... & {\it 0.154
0.034}\\
A2029 & 0.35 1070 ({\it 1137}) & 0.38  8.83 & 0.12 0.62 & 1.70& 
0.121 0.004 {\it (0.152)} & 0.158 0.006 {\it (0.185)} & 0.194 0.008 {\it (0.215)} & {\it 0.227 0.011}\\
A2142 & 0.50 1130 ({\it 1189}) & 0.61  8.57 & 0.19 0.60 & 1.79&
0.145 0.019 {\it (0.164)} & 0.190 0.025 {\it (0.208)} & 0.237 0.031 {\it (0.239)} & {\it 0.255 0.033}\\
A2163 & 1.60 1680$^{\dagger}$ ({\it 1448})& 1.09  9.16 & 0.36 0.65 & 1.96&
0.164 0.010 {\it (0.120)} & 0.194 0.012 {\it (0.154)} & 0.203 0.014 {\it (0.172)} & {\it 0.183 0.014}\\
A2204 & 0.30 1150 ({\it 1183}) & 0.39  9.25 & 0.12 0.66 & 1.65&
0.117 0.029 {\it (0.137)} & 0.151 0.037 {\it (0.160)} & 0.162 0.040 {\it
(0.179)} & {\it 0.185 0.046}\\
A2218 & 0.70 1100 ({\it 1039}) & 0.99 10.32 & 0.30 0.70 & 1.52& 
0.125 0.005 {\it (0.135)} & 0.157 0.007 {\it (0.156)} & 0.184 0.009 {\it (0.162)} & {\it 0.162 0.008}\\
A2219 & 1.30 1600 ({\it 1373}) & 1.59 11.51 & 0.48 0.79 & 1.97& 
0.142 0.009 {\it (0.130)} & 0.171 0.011 {\it (0.154)} & 0.180 0.011 {\it (0.157)} & {\it 0.155 0.011}\\
A2244 & 0.35 920 ({\it 1039}) & 0.39  8.40 & 0.11 0.59 & 1.50&
0.123 0.063 {\it (0.133)} & 0.163 0.083 {\it (0.171)} & ... & {\it 0.204
0.104} \\
A2256 & 1.60 1150 ({\it 1039}) & 2.34 13.60 & 0.56 0.82 & 1.87&
0.151 0.019 {\it (0.173)} & 0.209 0.024 {\it (0.204)} & 0.216 0.025 {\it (0.202)} & {\it 0.196 0.023}\\ 
A2319 & 1.00 1200 ({\it 1189}) & 1.06  8.77 & 0.35 0.62 & 1.90&  
0.148 0.006 {\it (0.150)} & 0.195 0.008 {\it (0.199)} & 0.227 0.011 {\it (0.226)} & {\it 0.242 0.013}\\
A2390 & 0.65 1300 ({\it 1299}) & 0.64  9.25 & 0.24 0.67 & 1.70&
0.151 0.021 {\it (0.126)} & 0.175 0.024 {\it (0.150)} & 0.202 0.029 {\it (0.164)} & {\it 0.169 0.024}\\
A2507 & 1.50 1240 ({\it 1195}) & 2.63 12.53 & 0.69 0.79 & 1.63&
0.088 0.023 {\it (0.081)} & 0.133 0.035 {\it (0.105)} & 0.140 0.038 {\it (0.111)} & {\it 0.111 0.030}\\
A2744 & 3.80 1950$^{\dagger}$ ({\it 1293}) & 2.76 15.21 & 0.62 0.86 &1.71&
0.162 0.013 {\it (0.121)} & 0.187 0.014 {\it (0.141)} & ...  & {\it 0.132
0.010} \\
A3112 & 0.23 700 ({\it 790}) & 0.26  8.68 & 0.08 0.60 & 1.17&
0.171 0.088 {\it (0.200)} & 0.230 0.118 {\it (0.258)} & ...  & {\it 0.277 0.143} \\
A3266 & 1.40 1200 ({\it 1103}) & 2.83 13.77 & 0.86 1.01 & 1.93&
0.111 0.007 {\it (0.160)} & 0.185 0.011 {\it (0.199)} & 0.204 0.012 {\it (0.204)} & {\it 0.199 0.013}\\
A3888 & 0.80 1280 ({\it 1170}) & 0.68  9.35 & 0.24 0.67 & 1.66&
0.123 0.025 {\it (0.129)} & 0.145 0.029 {\it (0.153)} & ... & {\it 0.170
0.035}\\
IRAS 09104 & 0.06 1100 ({\it 1137}) & 0.18 10.09 & 0.09 0.69 & 1.12&
0.063 0.038 {\it (0.089)} & 0.082 0.049 {\it (0.102)} & ... & {\it 0.106 0.064}\\ 
MS 1358 & 0.40 1100 ({\it 1068}) & 1.48 14.29 & 0.35 0.87 & 1.45& 
0.088 0.075 {\it (0.097)} & 0.117 0.101 {\it (0.095)} & ... & {\it 0.086 0.074}\\
MS 2137 & 0.25  930 ({\it 889}) & 0.18 11.48 & 0.09 0.81 & 1.08&
0.153 0.049 {\it (0.191}) & 0.150 0.048 {\it (0.188)} & ... & {\it 0.189
0.061}\\
PKS 0745 & 0.45 1200 ({\it 1150}) & 0.36  9.25 & 0.11 0.65 & 1.68& 
0.143 0.025 {\it (0.115)} & 0.158 0.028 {\it (0.135)} & 0.179 0.032 {\it (0.153)} & {\it 0.160 0.029}\\
Triang. Aus. & 1.00 1300 ({\it 1239}) & 1.41 10.40 & 0.48 0.75 & 2.15&
0.115 0.012 {\it (0.126)} & 0.156 0.016 {\it (0.155)} & 0.172 0.018 {\it (0.165)} & {\it 0.169 0.020}\\
Zw 3146 & 0.20 1310 ({\it 1311}) & 0.19 10.26 & 0.07 0.74 & 1.47&
0.094 0.044 {\it (0.125)} & 0.106 0.049 {\it (0.140)} & ... & {\it 0.159
0.074}\\
\hline
\end{tabular}
\end{table*}


The standard primordial nucleosynthesis theory indicates the present
universal ratio of baryons to photons as the free parameter to be
constrained through the observations of the abundances of the
light-elements. Once the microwave background temperature is fixed, and
models with three light neutrinos adopted, one can estimate the baryonic
mass density in units of the critical density, $\Omega_{\rm b}= \rho_{\rm
b} / \rho_{\rm c}$. Recently, conflicting estimates of the abundance of
deuterium, D, have raised questions on the robustness of a general value
for $\Omega_{\rm b}$ (e.g. Hogan 1997).
A high D abundance agrees with $^4$He and $^7$Li
measurements and with the baryonic cosmic budget (Fukugita et al. 1998)
and constrains $\Omega_{\rm b} h^2_{50}$ between 0.020 and 0.064 (with a
central value of about 0.04; e.g. Songaila, Wampler \& Cowie 1997), 
on the other hand a low D/H ratio match Galactic chemical evolution 
and local measurements better.  In this case, $\Omega_{\rm b} h^2_{50}$ is
$0.076 \pm 0.004$ (cf. Burles \& Tytler 1997). 

Thus, if regions that collapse to form rich clusters in an Einstein-de 
Sitter Universe retain the same value of $\Omega_{\rm b}$ as the rest of
the Universe, only a few per cent of cluster masses can be due to
baryons (mostly gas in the ICM, but also stars in galaxies and eventual
dark and cool baryons), in opposition with the observed 10--30 per cent
(Briel et al. 1992, White et al. 1993, White \& Fabian 1995, David, Jones 
\& Forman 1995, this paper). 

It is worth noting that, historically, X-ray 
observations have always shown a relatively high baryon fraction 
in clusters (e.g. Stewart et al. 1984); White \& Frenk (1991)
highlighted the discrepancy for the Coma cluster when new tighter 
and lower constraints from nucleosynthesis were published by Walker et 
al. (1991). 

In our case, the gas fraction, $f_{\rm gas}$, is estimated 
by the ratio of the gas mass determined in the deprojection (and
fitting) analysis to the gravitational mass 
estimated through the hydrostatic equation.
The measured values of $f_{\rm gas}$ at 0.5, 1, 1.5 Mpc (where
available, i.e. for radii less than $R_{\rm out}$) and $r_{500}$ (from the 
best-fit parameters of the NFW gas density profile) are quoted in Table~2.


All the values quoted at $r_{500}$ are obtained using the best fit
parameters as described in Sect.~3.1. 
The 68 per cent uncertainty on $f_{\rm gas}$ is obtained by propagating 
the errors on the gas mass and total mass. The former are
obtained in the deprojection analysis, by perturbing the surface brightness
profile 100 times, according to the Poisson error on the counts in each
radial bin. 
The uncertainty on the gravitating mass comes from assuming the above
dependence upon the gas temperature, $M_{\rm tot} \propto T_{\rm
gas}^{3/2}$, and propagating the relative error on the temperature itself,
i.e. $\epsilon(M_{\rm tot}) / M_{\rm tot} = 1.5 \epsilon(T_{\rm
gas})/T_{\rm gas}$.

To quantify the deviation between the deprojected and fitted results,
we calculate at $R_{\rm out}$ for each cluster the quantity \( (f_{\rm
gas, DEPROJ} - f_{\rm gas, FIT})/ \epsilon(f_{\rm gas}) \). 
On the sample of 36 clusters, we measure a median deviation of $+0.48
\epsilon(f_{\rm gas})$, with a range of $-1.93,+2.81$. The largest
deviation is due to A2744, which shows a minor merger in the X-ray
image and strong disagreement between the X-ray temperature and optical
dispersion when isothermality is assumed.

\subsection{Comparison with previous work}

Our sample of clusters with high X-ray luminosity contains several
clusters already analyzed, with the {\it Einstein} and {\it ROSAT} HRIs. 
Only 6 of them have been studied also with the PSPC (i.e. A401, A1795,
A2029, A2163, A2256, A2390). 

We recall that the use of PSPC profiles extracted to about 2 Mpc (i) does
not allow a good resolution in the inner part of the clusters, both in the
deprojection analysis and in the fitting procedure, where we also cut the
profile below 0.2 Mpc (or $0.1 \ r_{500}$) to avoid the contribution from
any cooling flow, and
(ii) mainly weights the outskirts in the fitting analysis (just a few bins
are located in the core). These points have to be borne in mind when we
compare our results on the gas fraction measured in the inner part (e.g
500 kpc) with estimates obtained through observations which are more
efficient for that purpose (cf. the results from HRI observations in Allen
\& Fabian 1998).

In Fig.~\ref{fgas6}, we plot the difference in standard deviations between
previous estimates and our results from the fitting procedure.  We
generally underestimate the gas fraction in the core of cooling flow
clusters (e.g. A478, A1689, A1835, A2029, A2142) when compared to the
Allen \& Fabian (1998) results.

In A1795, the effect of the cooling flows produces a remarkable mismatch
between our estimates and the results of David et al. (1995; about $3.7
\sigma$) and Allen \& Fabian (1998; $\sim 5.5 \sigma$). 
On the other hand, the conclusion from Briel \& Henry (1996) shows an
opposite trend with respect to these and to our result from the fitting
procedure, while White et al. (1997) are in good agreement with our
value at 1 Mpc. 

\subsection{The distribution of $f_{\rm gas}$ and its dependence upon
redshift}

We select from our sample the clusters where the value of $f_{\rm gas}
(<r_{500})$ is twice its  uncertainty, i.e. we apply the selection
criterion that $f_{\rm gas} / \epsilon(f_{\rm gas}) > 2$
(Fig.~\ref{fgas6bis}).
This leaves 30 clusters (A545, A586, A2244, A3112, IRAS09104, MS1358,
excluded) as our definitive sample. 
As discussed in Sect.~2, this sample is not complete in any sense.
In the following analysis, we adopt the null hypothesis that (i) the gas fraction
is constant with redshift and total mass, (ii) our high X-ray luminosity clusters are a
homogeneous sample where no contamination is expected by,
say, cooling flows. (Note that the gas temperatures are obtained from spectral 
analyses which exclude the cooling flow regions.)

The weighted mean of the values of $f_{\rm gas} (r_{500})$ is 
$0.176 (\pm 0.003)$, with a variance of 0.037. 

The mean, however, is statistically a poor estimator, being sensitive to 
(i) the replacement of even a small part of the data with
different values, (ii) the assumed underlying population, (iii) the
increase in the number of data to get better information. In this sense,
the median is certainly a better estimator. But, an even better (i.e.  
more resistant, robust and efficient) estimator is the biweight location
(Beers, Flynn \& Gebhardt 1990). Using this estimator,
we calculate $f_{\rm gas} (< r_{500}) = 0.171 \pm 0.035$. 
Moreover, we can use the clusters where a significant cooling flow
is present (cf. Table~1 in Allen \& Fabian 1998) to estimate the gas mass
fraction characteristic of more relaxed systems.
We select 13 clusters (A478, A1068, A1413,
A1689, A1795, A1835, A2029, A2142, A2204, A2390, MS2137, PKS0745, Zw3146)
and measure a biweight of $0.168\pm0.030$, that is consistent with the 
distribution in the whole sample.

Both the weighted mean and our estimate of the biweight location 
are completely consistent with the biweight estimate of the gas
fraction, at $r_{500}$, of $0.170 \pm 0.008$ quoted by Evrard (1997) and
calculated from two different published samples (White \& Fabian
1995, David et al. 1995) after revision of the total mass of the David et
al. clusters. 

We now verify if the independent measurements of $f_{\rm gas}$ are
compatible in a significant way.
Applying the $\chi^2$ test to our sample of $f_{\rm gas}$, we
find that these are {\it not} compatible at the $> 99.9$ per
cent confidence level when the weighted mean is adopted as 
the representative value. 
In fact, the deviations between our values of $f_{\rm gas}(r_{500})$
can be up to a factor of 3 (cf. A483 and A2142; see also
conclusions on A1060 and AWM7 in Loewenstein \& Mushotzky 1996).

Carrying over our robust approach to these statistical issues, we
investigate the compatibility between each value of $f_{\rm gas}$ and its
average representative estimate, following the considerations in Press
(1996) on the Bayesian combination of apparently incompatible measurements
of the underlying quantity (in our case, $f_{\rm gas}$). This method,
which weighs a weighted sum of `good' and `bad' Gaussians attributed to
each measurement with an a priori probability that the experiment is
`correct', provides a distribution of probability for the value under
investigation and a judgment on the goodness of each experiment. Applying
it, we find that all the measurements, with a significantly low
probability of less than 0.01 per cent that they are all wrong, are
consistent with a distribution, roughly symmetric, peaked at 0.168 and
with a dispersion of of $[-0.030, +0.036]$ (Fig.~\ref{fgas7}). 
This distribution is
completely consistent with the biweight location and scale. Hereafter, we
consider this value, with the respective ``dispersion'' around the
``mean'', as the representative estimate of $f_{\rm gas}$ in our sample.
In Fig.~\ref{fgas7}, we also indicate the estimated value of $\Omega_{\rm
b}$ from low D abundance. When we compare this value with the calculated
probability distribution of $f_{\rm gas}(r_{500})$, we locate it on the
wing of the distribution as very unlikely (probability of 0.7 per cent). 

We investigate now two other issues related to the distribution of gas in
clusters: (i) the dependence of the gas fraction on the radius within
each cluster, and (ii) its constancy with the cosmological time. 
These issues are strictly related to the formation and evolution of
clusters of galaxies in the present cosmological scenario.

The observed structure in the Universe results from the evolution
of gravitational instability. On scales between $10^{13}$ and
$10^{15} M_{\odot}$, gravity is assumed to be the only force driving the
formation of groups, poor clusters and clusters of galaxies. In this
scenario, where the evolution is an entirely self-similar process, the gas
and the total mass should have the same distribution. 

This is not what is generally observed. In clusters, data analysis (e.g. 
White \& Fabian 1995 and David, Jones \& Forman 1995) and hydrodynamics
simulations (Evrard 1997) show an increase of $f_{\rm gas}$ with radius. A
way to explain the discrepancy between the distribution of gas and the
underlying dark matter is to take into account physical phenomena able
to redistribute the energy in the cluster, like galactic winds,
ram-pressure stripping and heat input by supernovae type II (Evrard 1997,
Metzer \& Evrard 1998, Cavaliere, Menci \& Tozzi 1998, Wu, Fabian \&
Nulsen 1998). In particular, the role of galactic winds is now
well-studied in raising the gas entropy from the value achieved after
gravitational collapse and flattening the gas distribution inside the
cluster. This should affect both the global value of $f_{\rm gas}$ and its
radial dependence.

In our case, which selects high-luminosity clusters, the median value of
$T_{\rm gas}$ of 9 keV should save our local $f_{\rm gas}$ estimates
from dropping significantly with respect to the global value. 
Feedback from galaxy formation affects the low-temperature clusters
most strongly (e.g. Metzer \& Evrard 1998), where
the total intracluster thermal energy becomes comparable to the energy
input from feedback.

On the other hand, we can investigate the redistribution of 
cluster energy, describing the radial variation of the gas fraction. 
In Fig.~\ref{fgas8}, we show the general increase in $f_{\rm gas}$ when
calculated at 0.3 and 1.0 $r_{500}$. 
To quantify this trend, 
we collect in 14 radial bins equally spaced between [0.3, 1.0]
$r_{500}$ the averaged gas fraction (with an error on the mean obtained
from the propagation of the individual errors), making a composite $f_{\rm
gas}$ profile.
Then, fitting a power law to the 14 bins (Fig.~\ref{fgas9}), we obtain
$f_{\rm gas}(r) \propto (r/r_{500})^{0.20 \pm 0.02}$.
This result is slightly steeper than the estimate of about 0.13--0.17
from simulations (cf. Evrard 1997). 

Moreover, in the two clusters (A644, A1651) for which the radius
where a dark matter overdensity of 200 is surveyed (i.e., $r_{200} <
R_{\rm out}$), we observe that the gas fraction increases by about a
further 15 per cent from $r_{500}$ to $r_{200}$.
 
The evidence of a positive gradient underlines a lower concentration of
the gas with respect to the dark mass, that is consequence of both our
assumption on the dark matter profile (at $r\sim r_{500}$, $\rho_{\rm DM}
\propto r^{-2.4}$; cf. N-body simulations results, for example Thomas et
al. 1998) and our best-fit results on the slope of the gas density (i.e.
$\rho_{\rm gas} \propto r^{-3 \beta} \propto r^{-2.2}$, that implies a
dependence of $M_{\rm gas}$ on $r^{0.8}$).
In other words, converting the dependence of the gas fraction upon  
radius to a dependence on the different mass components, we conclude that 
$f_{\rm gas}(r) \propto [M_{\rm gas}(r)/M_{\rm gas}(r_{500})]^{1/4} 
\propto [M_{\rm tot}(r)/M_{500}]^{1/3}$.

Another argument against simple self-similar cluster evolution would be
a variation of $f_{\rm gas}$ with redshift. 

First, the use of Spearman's rank-order correlation shows that $f_{\rm
gas}$ is correlated with redshift at a confidence level $> 99$ per cent.
We investigate this further by dividing the sample according to the median 
value in redshift ($z$=0.1680) and calculating the biweight location (and 
respective bootstrap error) of the $f_{\rm gas} (r_{500})$ values. 
We obtain $0.192 (\pm 0.012)$ and
$0.159 (\pm 0.009)$, at low and high redshifts, respectively. 
The evidence of a higher local gas fraction is also clear from
Fig.~\ref{fgas10}, where we sample the data in 3 redshift bins of 10
elements each.  
Each bin is represented with the biweight location, and the respective
error. We fit then a power law, $f_{\rm gas, 0} \times (1+z)^{-\alpha}$, and
measure $f_{\rm gas, 0} = 0.215 (-0.019, +0.020)$ and $\alpha = 1.75
(-0.65, +0.65)$, where the errors are $1 \sigma$ deviation.
Again, the slight negative evolution in the gas fraction appears
significant at $\sim 2.7 \sigma$ level (estimated from the value of
$\alpha$ and its error). 

\begin{table}   \caption{The biweight values of $f_{\rm gas}$ (with
the respective bootstrap error) for different subsamples selected, first,
in redshift with respect to the median value of 0.1680, and then in
$M_{500}$ (the median values for the low and high redshift sample are 0.97
and 1.08 $\times 10^{15} M_{\odot}$, respectively). In parenthesis, the
number of values for each subsample is quoted. }
\begin{tabular}{| l c c |} \hline
      & low $z$  &  high $z$  \\
low $M_{500}$  & $0.189 \pm0.020$ (7)  & $0.167 \pm0.012$ (7) \\
high $M_{500}$ & $0.192 \pm0.017$ (8)  & $0.151 \pm0.012$ (8) \\
 \hline
\end{tabular}   
\end{table}
 
We have also looked for trends of $f_{\rm gas}$ with the total mass,
$M_{500}$, (Fig.~\ref{fgas11}) considering the biweight gas fraction
values calculated in subsamples that were selected according to the median
of the redshifts and of the masses (Table~3).
In particular, we search for the minimum in the $\chi^2$ distribution,
when the data in Table~3 are compared with the model 
\begin{equation}
f_{\rm gas}(r_{500}) = {\rm const} \times M_{500}^{-\alpha_0} 
(1+z)^{-\alpha_1}. 
\label{fgas:eq6}
\end{equation}
We obtain a minimum $\chi^2$ of 0.39 (one d.o.f.)
with the best-fit parameters: $\alpha_0 = 0.15$, $\alpha_1 = 1.24$. 
In Fig.~\ref{fgas12}, we plot the 68 and 95 per cent errors contour for
the two slope parameters. The dependence on the redshift is well in
agreement with the previous results plotted in Fig.~\ref{fgas10}. 
On the other hand, there is no evidence for any statistically significant
dependence of $f_{\rm gas}(r_{500})$ on $M_{500}$.


Allen \& Fabian (1998) show a decrease in the gas
fraction of clusters at higher temperature (cf. their Fig.~3). If we
replace $M_{500}$ in equation~\ref{fgas:eq6} with the
intracluster temperature, we find that the slope of $T_{\rm gas}$ is
within the range [-1.4 ,0.4] at the 95 per cent confidence level. But
putting $\alpha_1 = 0$, $\alpha_0$ has to be larger than 0 (95 per cent
c.l.), in agreement with Allen \& Fabian results.

We also note that, from the scaling low between mass and temperature ($M
\propto T_{\rm gas}^{3/2}$), any apparent dependence of $f_{\rm gas}$ upon
$T_{\rm gas}$ becomes weaker by a factor 2/3 when applied to mass.

This scenario implies that the gas component in the X-ray highly-luminous
systems considered here is almost independent of the mass and only 
slightly on the temperature, once the redshift dependence is taken into
account.

Any apparent decrease with redshift of the gas fraction, however, has been
recently questioned. 
Following an original idea of Sasaki (1996), Cooray (1998) and Danos \&
Pen (1998; see also Rines et al. 1998) have shown the
angular distance$-$redshift relation, $d_{\rm ang}(z,q_0)$, that we write
in equation~\ref{fgas:eq3}, to be the major factor responsible for the
apparent negative evolution.
From the definition of the gas fraction, it holds that $f_{\rm
gas} \propto d_{\rm ang}^{3/2}$ and, consequently, $f_{\rm gas}$ tends to
be lower at higher redshifts in a high density universe. 

We consider the changes in $f_{\rm gas}$ for cosmologically 
different scenarios in Fig.~\ref{fgas10}.  We lower
$\Omega_{\rm 0, m}$ to 0.2, applying both an open universe and a flat
one with a cosmological constant $\Omega_{\Lambda} = 1 - \Omega_{\rm 0,
m}$ [cf. eqn.~(25)  in Carroll, Press \& Turner 1992]. 
In the latter case the change appears more significant in flattening 
$f_{\rm gas}$ to a non-evolving value. Fitting a power-law to 
these  values, we obtain $f_{\rm gas} = 0.210 (-0.020, +0.022) \times
(1+z)^{-0.61 (-0.66, +0.64)}$ with a reduced $\chi^2$ of 1.3. 
All the errors are one standard deviation. 

Also when we apply equation~\ref{fgas:eq6} to the case $[\Omega_{\rm 0,
m}, \Omega_{\Lambda}] = [0.2, 0.8]$, that shows the better agreement
with the constant $f_{\rm gas}$ assumption, we observe a significant
flattening in the redshift dependence.
The best-fit parameters ($\chi^2 = 0.22$ for 1 d.o.f.)  are now
$(\alpha_0, \alpha_1) \sim (-0.1, 0.5)$, with a range for
the $\alpha_0$ values of $[-0.5, 0.7]$ at the 95 per cent confidence level
(dotted contours in Fig.~\ref{fgas12}). 


Thus, a low density Universe not only matches the observed gas fraction
with the baryonic amount provided during the primordial nucleosynthesis,
but also flattens to a constant (in look--back time) gas mass fraction
of about 0.196 (biweight estimate).




\subsection{The dependence of $f_{\rm gas}$ on the temperature profile}

\begin{figure}
\psfig{figure= 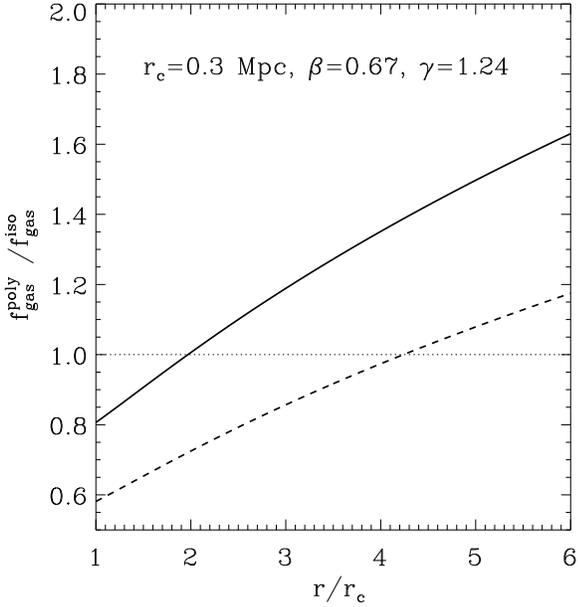,width=.5\textwidth,angle=0}
\caption{The gas fraction relative to the isothermal case, after 
correction for the gradient in the temperature parametrised by the
polytropic equation (see Appendix). When we assume the isothermal
gas temperature equal to the maximum value of the temperatures 
allowed from the polytropic profile (i.e. the central value, $T_0$), 
we observe an increase of the polytropic $f_{\rm gas}$
estimate up to 60 per cent with respect to the isothermal $f_{\rm gas}$
value at $r \sim 6 r_{\rm c}$ (solid line). 
Using an emission-weighted average of the temperatures spanned from the
polytropic profile
as reference value for the isothermal profile (instead of the central
value), the correction on $f_{\rm
gas}$ is by about 17 per cent at the same radius (dashed line). 
} \label{fgas14} \end{figure}

In this Section, we remark on the role of a negative
gas temperature gradient (instead of an isothermal profile) in
estimating the cluster gas fraction.
Markevitch et al. (1998) have recently claimed, from the analysis
of a sample of clusters observed from the X-ray satellite {\it ASCA},
that a negative temperature gradient is generally present and well
represented by a polytropic function with index, $\gamma$, of 1.2-1.3
between 1 and 6 X-ray core radii. They have assumed a
$\beta$-model with $[\beta, r_{\rm c}]$ of [0.67, 0.3 Mpc] to
constrain the gas density, and then used the polytropic equation to fit
the temperature profiles.

In the Appendix, we derive the equations necessary to describe the
polytropic state of the intracluster plasma. Using the results of
Markevitch and collaborators, we make a correction to the total mass via
equation~\ref{fgas:a8} of about 1.2 -- 0.6 over the radial range $[ 1, 6]
r_{\rm c}$ on the total mass. It yields a larger gas fraction with respect
to the isothermal case, above 2 $r_{\rm c}$.  At $r \sim 6 \ r_{\rm c}
\sim r_{500}$, the polytropic value of $f_{\rm gas}$ can be larger by a factor
of 1.6 than the isothermal estimate (Fig.~\ref{fgas14}).

We have also fitted eqn.~\ref{fgas:a5}--\ref{fgas:a6} to the profiles of
the 34 clusters in our sample with enough data points, to obtain $\gamma =
1.25 \pm 0.13$.  None of these polytropic fits, however, is better than
the isothermal one at 95 per cent confidence level using the F-test. 

We note that steep electron temperature profiles need not be
representative of the state of the gas. As shown by, e.g., Ettori
\& Fabian (1997) and Takizawa (1998), the efficiency of the Coulomb
collisions between ions and electrons has to be considered when a such
negative gradient is observed in clusters, even if they underwent an
ancient merger. Thus, when the equipartition time is comparable to the age
of the last, large, merging event, it is also true that the mean gas
temperature profile {\it is generally flatter} than the (emission-weighted)
electron temperature profile estimated from X-ray observations.

In conclusion, the presence of a non-isothermal temperature profile tends
to increase the gas fraction value at $r_{500}$, more or less
significantly depending on our underestimate of the drop of the
temperature in the outskirts of each cluster with respect to the assumed
representative isothermal value. 
Whether this is a systematic or random uncertainty in our results depends
upon whether all clusters have similar profiles or not up to $r_{500}$. 
Current data are unable to clarify this important issue.

\section{$\Omega_{\rm 0, m}$ from the gas fraction}


The range of observed cluster gas fractions can now be compared with the
results of primordial nucleosynthesis calculations. Assuming that the
current best estimate for $H_0 = 73 \pm 14$ km s$^{-1}$ Mpc$^{-1}$ (Freedman
et al. 1998), and the constraint
\begin{equation}
\Omega_{0, \rm m} < \ \frac{\Omega_{\rm b}}{f_{\rm gas}} \left( \frac{H_0}
{50 \ \mbox{km s$^{-1}$ Mpc$^{-1}$} } \right)^{-0.5},
\label{fgas:eq7}
\end{equation}
we obtain $\Omega_{0, \rm m} < 0.20 (-0.13, +0.13)$, adopting
$\Omega_{\rm b} = 0.04$ which corresponds to the high deuterium abundance 
estimate, and $\Omega_{0, \rm m} < 0.37 (-0.16, +0.09)$, using
$\Omega_{\rm b} = 0.076$. The error bars come from the propagation of
the accepted range for the given $\Omega_{\rm b}$ (see beginning
of Sect.~4). From the latter estimate, we can put an upper limit on the
density of the Universe as inferred from the measured gas fraction in our
sample of highly-luminous clusters: $\Omega_{0, \rm m}$ has to be
less than 0.56 at the 95 per cent confidence level.
We note that our estimate for $\Omega_{\rm 0,m}$ is conservative due to our
use of an `average' value for $f_{\rm gas}$. If we adopt the 95 per cent
lower limit to the highest reliable values in our sample of gas fraction 
(considering 10 per cent of the sample, that of A2142, A2319, A401,
A2029) to constrain the cosmological parameter, we require that
$\Omega_{\rm 0,m}<0.34$. 

 
Using the conservative constraint, it is straightforward to require
$\Omega_{\Lambda} = 1 - \Omega_{0, \rm m} > 0.44$ (95 per cent confidence
level), still marginally consistent with the recent results on the
magnitude-redshift relation for the type Ia supernovae (Perlmutter et al. 
1998). 


We have neglected so far any other baryonic contribution, like galaxies
and baryonic dark matter. The luminous mass in galaxies is about a fifth
of the gas fraction, i.e. $g= f_{\rm gal}/f_{\rm gas} \sim 0.2
h_{50}^{1.5}$ (White et al. 1993, Fukugita et al. 1998).
The cluster baryonic dark matter is suggested to be in the form of cold
clouds or low mass stars and brown dwarfs deposited by cooling flows
(Thomas \& Fabian 1990, Fabian 1994).
Also on galactic scale, low mass dark objects can be deposited by a
cooling flow in the halo during first collapses of a protogalaxies
(Nulsen \& Fabian 1997). 

The last equation can be then rewritten considering these other
baryonic contributions. Defining $b$ as the ratio of this baryonic dark
matter to the gas fraction, and assuming it is independent of $H_0$, we
can
write:
\begin{equation}
\Omega_{0, \rm m} = \ \frac{\Omega_{\rm b}}{f_{\rm gas}} \ 
\frac{h_{50}^{-0.5}}{ 1 + (b+g) h_{50}^{1.5} }.
\label{fgas:eq8}
\end{equation}

\begin{figure}
\psfig{figure= 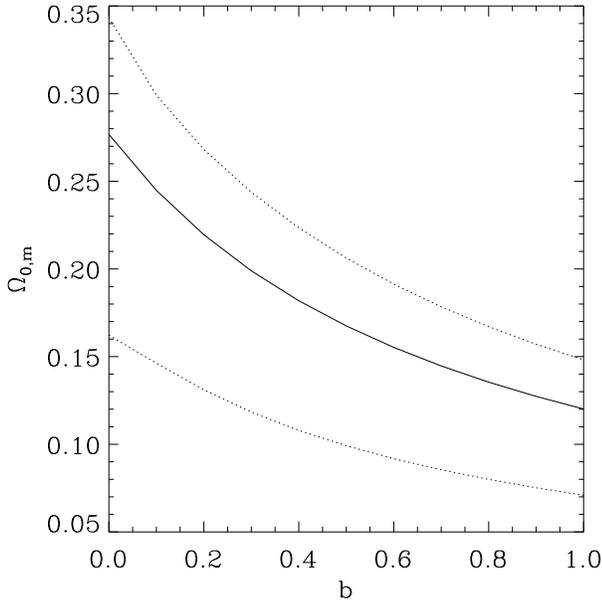,width=0.5\textwidth,angle=0}
\caption{Expectation values for $\Omega_{0, \rm m}$ when a baryonic
contribution is considered in addition to intracluster gas and stars in
galaxies [cf. equation~\ref{fgas:eq8}].  The case $b=0$ corresponds to the
$\Omega_{0, \rm m}$ value estimated considering only $f_{\rm gas}$ and
assuming a low value for D/H.  
} \label{fgas13} \end{figure}

For different $b$ values, using the best estimates from our previous
analysis and propagating the errors as usual, we plot in
Fig.~\ref{fgas13} the estimated $\Omega_{0, \rm m}$ with $1 \sigma$
uncertainty. 
For example, recent estimate on the universal primordial baryon fraction from
MACHO results (Steigman \& Tkachev 1998), assumed baryonic and in the halo of
the Galaxy, give us a value of $b \sim 0.6$, with a consequent $\Omega_{0,
\rm m}$ of about 0.15.

Here we note that the contribution from baryonic dark matter can be 
about 3 times the luminous matter in normal stars, as predicted from
standard model of gas processes in galactic formation (Nulsen \& Fabian
1997).
This can partially explain the large scatter in the gas fraction
distribution (for example, cf. Fig.~\ref{fgas11}), suggesting 
that most of the baryons in clusters with lower values of $f_{\rm gas}$
are in the form of dark matter.  


\section{SUMMARY AND CONCLUSIONS}

In our sample of 30 hot (median $T_{\rm gas} = 9$ keV) and intermediate
distance (median $z = 0.1680$) clusters of galaxies, we have
applied both the deprojection technique and the fitting approach, using
both (a) the $\beta$-model and (b) a gas profile obtained from the
Navarro-Frenk-White potential. For (a) we find that the median best-fit
parameters are $[\beta, r_{\rm c}] = [0.71, 0.24 \ {\rm
Mpc}]$; for (b), $[\eta, r_{\rm s}] = [10.20, 0.75 \ {\rm Mpc}]$.

The two techniques produce the same values of $f_{\rm gas}$ within a
median deviation of less than $1 \sigma$.

Assuming an isothermal profile for the gas temperature, with the caveats
discussed in Sect.~4.3 on the role played by a decrease of the
temperature outward, we obtain the following results on the
gas fraction, $f_{\rm gas}$:

\begin{enumerate}
\item At the radius where the cluster overdensity is 500, the measured
values of $f_{\rm gas}$ have a biweight location and scale
of $0.171\pm0.035$. When Bayesian statistics are adopted, the probability
distribution is peaked at 0.168 and has a 95 per cent range of
[0.101, 0.245] (Fig.~\ref{fgas7}).

With respect to this distribution, the highest estimate of $\Omega_{\rm
b}$ (corresponding to a low deuterium abundance) has a probability of
$7.2 \times 10^{-3}$ in an Einstein-de Sitter Universe 
($\Omega_{0, \rm m} = 1$).

When we consider the more relaxed clusters with a central cooling flow, 
we measure $0.168 \pm 0.030$.

If we consider the cosmological correction in the angular size--redshift 
relation for a flat Universe with $\Omega_{\Lambda} = 0.8$,
we measure the biweight values of $0.196 \pm 0.035$.

\item Many of the individual estimates of $f_{\rm gas}$ are inconsistent
with the average value or with each other, confirming that there are real
differences in the measured gas fractions in clusters
(Fig.~\ref{fgas6bis}, \ref{fgas8}, \ref{fgas11}).  
It remains possible that some of this spread is due to a breakdown in our
assumptions. There may, for example, be a merger taking place along the
line of sight in some clusters. If the gas is actually more extended along
the line-of-sight then we have underestimated the gas fraction,
conversely if it is foreshortened then the gas fraction is overestimated.

\item The average dependence upon the radius within a cluster is $r^s$,
with $s \sim 0.20$ (Fig.~\ref{fgas9}). 
This follows from our best-fit results on the slope of the gas density,
$\rho_{\rm gas} \propto r^{-2.2}$, and our assumption of the
Navarro-Frenk-White functional form for the dark matter distribution. 
In general, we can write $f_{\rm gas} \propto r^{-2.2 -s_{\rm DM}}$, 
where $s_{\rm DM}$ is the slope of the dark matter profile. 

\item In our sample of high-luminosity clusters, there is a highly
significant correlation between $f_{\rm gas}(r_{500})$ and redshift, that
we quantify fitting the function $f_{\rm gas,0} \times (1+z)^{-\alpha}$:
$f_{\rm gas,0} = 0.215^{+0.020}_{-0.019}$, $\alpha = 1.75 \pm 0.65$ 
($\Omega_{0, \rm m} = 1$). When we take it into account,
there is then only a mildly significant dependence of $f_{\rm
gas}(r_{500})$ upon the temperature (and, even weaker, upon the mass).
The dependence upon redshift weakens (i.e. the significance of any
decrease in $f_{\rm gas}$ is reduced) in a low matter density universe, in
particular if a cosmological constant is present. 
The normalization of the power law, i.e. the extrapolated local value of the
gas fraction, $f_{\rm gas,0}$, remains stable at 0.21. 


\item Adopting both a low and a high abundance for deuterium and requiring
that any physical source in redistributing the energy within the cluster
cannot produce large variations in the value of $f_{\rm gas}$ quoted
above [cf. (i)], we constrain $\Omega_{0, \rm m} < 0.56$ at the 95 per cent
confidence level. If we take the highest significant estimates of $f_{\rm
gas}$ in our cluster sample, we find that $\Omega_{0, \rm m} < 0.34$.

\item Future X-ray missions can constrain better the temperature
gradient in the electron (and electron+ion) population. If the indication of 
a negative gradient in $T_{\rm gas}$ is confirmed, $f_{\rm gas}$ rises
and $\Omega_{0, \rm m}$ drops by a factor that can be, at maximum, 1.6.
Such a large value of
$f_{\rm gas}$ could become a real concern for the physics of the formation
and evolution of clusters of galaxies in present cosmological scenarios.

\end{enumerate}

\section*{ACKNOWLEDGEMENTS} We are grateful to Steve Allen, Kelvin Wu,
David White and Vince Eke for useful discussions.
We thank the anonymous referee for comments which improved this work.
SE acknowledges support from PPARC, Cambridge European Trust and {\it
Alfred Toepfer Stiftung F.V.S.}, ACF the support of the
Royal Society.  This research has made use of data obtained through the
High Energy Astrophysics Science Archive Research Center Online Service,
provided by the NASA-Goddard Space Flight Centre.

\appendix
\section{Analytic models for the state of the intracluster gas}
The  Navarro, Frenk \& White (NFW) density profile, $\rho_{\rm NFW} = \rho_{\rm s}
x^{-1} (1+x)^{-2}$, generates a gravitational potential of the form:
\begin{equation}
\frac{d\phi}{dr}= 4\pi G r_{\rm s} \rho_{\rm s} \frac{\ln (1+x) -x/(1+x)}
{x^2},
\label{fgas:a1}
\end{equation}
where, defining $r_{\rm s}$ as a {\it  scale} radius, $x= r/r_{\rm s}$ and
$\rho_{\rm s} = \rho_{\rm c} \delta_{\rm c}
(1+z)^3 \Omega_0/\Omega_{\rm z} $, with $\delta_{\rm c}$ equal to the
characteristic density of the cluster and $\rho_{\rm c}$ to the critical
density (see Appendix in Navarro, Frenk \& White 1997).

Putting it in the hydrostatic equation, and assuming the isothermality for
the intracluster gas, $T_{\rm gas}(r) = T_0$,
\begin{equation}
\frac{1}{\rho_{\rm gas}}\frac{d\rho_{\rm gas}}{dr} = -\frac{\mu m_{\rm p}}
{kT_{\rm gas}} \frac{d\phi}{dr},
\label{fgas:a2}
\end{equation} 
we obtain
\begin{equation}
\rho_{\rm gas} = \rho_0 e^{-\eta} (1+x)^{\eta/x} = a_0 (1+x)^{\eta/x},
\label{fgas:a3}
\end{equation}
where 
\begin{equation}
\eta = \frac{4\pi G \rho_{\rm s} r^2_{\rm s} \mu
m_{\rm p}}{kT_0} = \frac{1.5 \delta_{\rm c} (1+z)^3 (\Omega_0/\Omega_{\rm z})
H_0^2 r^2_{\rm s} \mu m_{\rm p}}{kT_0}.
\label{fgas:a3bis}
\end{equation} 
(we define, as
usual, $G$ as the Gravitational constant, $\Omega_0$ and $\Omega_{\rm z}$
as the cosmological parameters at the present time and at redshift $z$,
and $H_0$ as the present-time Hubble constant). The gas density profile
is then integrated and fitted to the observed surface brightness profile
to constrain the free parameters $\rho_0, r_{\rm s}$ and $\eta$.
Hence, given $T_0$ and the best-fit parameters $r_{\rm s}$ and $\eta$,
$\delta_{\rm c} = \overline{\delta}/3 \times c^3/[\ln(1+c) - c/(1+c) ]$ and the
concentration parameter $c$ can be estimated from eqn.~\ref{fgas:a3bis}. 
Here, $\overline{\delta}$ is the mean interior overdensity that defines a
cluster over the background. 
It is generally assumed equal to about 200, that is the density contrast of a
virialized object in the nonlinear regime of spherical collapse.

The radius at which this mean overdensity is reached, $r_{\overline{\delta}}$,
is then defined $r_{\overline{\delta}} = c \ r_{\rm s}$ (Navarro, Frenk \& White
1995, 1997). NFW (1997) show that for a cluster identified at redshift $z$, 
the cosmological dependence of $r_{\overline{\delta}}$ is $(\Omega_0/\Omega_{\rm
z})^{-1/3} (1+z)^{-1} = (1+\Omega_0 z)^{-1/3} (1+z)^{-2/3}$. 
This decreases with the redshift and increases slightly with the lowering of the
density parameter: for example, at redshift 0.4, $r_{\overline{\delta}}$ is
lower than the local estimate by about 29 and 23 per cent for $\Omega_0$ equal
to 1 and 0.3, respectively.

We can also generalize to an intracluster gas with a polytropic equation of state
\begin{equation}
\rho_{\rm gas} = \rho_0 \left[ \frac{T_{\rm gas}(r)}{T_0}
\right]^{\frac{1}{\gamma -1}}
\label{fgas:a6}
\end{equation}
the consequences of the hydrostatic equilibrium with a cluster potential
described
by the NFW functional form (eqn.~\ref{fgas:a1}):
\begin{equation} 
\frac{d(kT_{\rm gas})}{dr} = \mu m_{\rm p} \frac{\gamma-1}{\gamma} \left( 
-\frac{d\phi}{dr} \right),
\label{fgas:a4}
\end{equation}
that integrated over the radial range of interest provides the equation:
\begin{equation}
\frac{T_{\rm gas}(r)}{T_0} = 1 + \eta \frac{\gamma-1}{\gamma} \left[ 
\frac{\ln (1+x)}{x} -1 \right], 
\label{fgas:a5}
\end{equation}
where $T_{\rm gas}(0) = T_0$ and $\eta = \eta(T_0)$.

The gas density profile is then obtained using equation~\ref{fgas:a6}.

The presence of a temperature gradient also affects the estimate of the total
gravitating mass, $M_{\rm tot}$, for which the gravitational potential is
unkwown: 
\begin{equation}
M_{\rm tot} = -\frac{kT_0 \ r^2}{\mu m_{\rm p} G} \ \gamma \left(
\frac{\rho_{\rm gas}}{\rho_0} \right)^{\gamma-2} \frac{d}{dr}
\left(\frac{\rho_{\rm gas}}{\rho_0} \right).
\label{fgas:a7}
\end{equation}
The case $\gamma =1 $ provides the usual equation to estimate the total
gravitating mass when the gas is isothermal.
In particular, the ratio between the polytropic and isothermal mass
estimates will be given by 
\begin{equation}
\frac{ M^{\rm poly}}{ M^{\rm iso}} = \gamma \left( \frac{\rho_{\rm
gas}}{\rho_0} \right)^{\gamma-1} = \gamma \frac{T_{\rm gas}(r)}{T_0}.
\label{fgas:a8}
\end{equation}

This equation has been used to estimate the correction on $f_{\rm gas}$
for the presence of a temperature gradient as plotted in Fig.~14. 



\begin{thebibliography}{} 
\bibitem[]{} Allen S.W., Fabian A.C., 1998, MNRAS, 297, L57
\bibitem[]{} Arnaud K.A., 1996, "Astronomical Data Analysis Software and
Systems V", eds. Jacoby G. and Barnes J., ASP Conf. Series vol. 101, 17
\bibitem[]{} Beers T.C., Flynn K., Gebhardt K., 1990, AJ, 100, 32
\bibitem[]{} Briel U.G., Henry J.P., B\"ohringer H., 1992, A\&A, 259,
L31
\bibitem[]{} Briel U.G., Henry J.P., 1996, ApJ, 472, 131
\bibitem[]{} B\"ohringer H., Tanaka Y., Mushotzky R.F., Ikebe Y.,
Hattori M., 1998, A\&A, 334, 789
\bibitem[]{} Buote D.A., Canizares C.R., 1996, ApJ, 457, 565
\bibitem[]{} Burles S., Tytler D., 1998, Space Science Reviews, 84 (1/2), 65 
\bibitem[]{} Carroll S.M., Press W.H., Turner E.L., 1992, ARAA, 30, 499
\bibitem[]{} Cavaliere A., Fusco-Femiano R., 1976, A\&A, 49, 137
\bibitem[]{} Cavaliere A., Menci N., Tozzi, P., 1998, ApJ, 501, 493
\bibitem[]{} Cooray A.R., 1998, A\&A, 333, L71
\bibitem[]{} Danos R., Pen U., 1998, astro-ph/9803058
\bibitem[]{} David L.P., Slyz A., Jones C., Forman W., Vrtilek S.D.,
\& Arnaud K.A., 1993, ApJ, 412, 479
\bibitem[]{} David L.P., Jones C., Forman W., 1995, ApJ, 445, 578
\bibitem[]{} Davis D.S., White III R.E., 1998, ApJ, 492, 57
\bibitem[]{} de Vaucoulers G., 1948, Ann. Astrophys., 11, 247 
\bibitem[]{} Durret F., Forman W., Gerbal D., Jones C., Vikhlinin
A., 1998, A\&A, 335, 41
\bibitem[]{} Ebeling H., Voges W., B\"ohringer H., Edge A.C., 
Huchra J.P., Briel U.G., 1996, MNRAS, 281, 799 
\bibitem[]{} Elbaz D., Arnaud M., B\"ohringer H., 1995, A\&A, 293,
337
\bibitem[]{} Ettori S., Fabian A.C., White D.A., 1997, MNRAS, 289, 787
\bibitem[]{} Ettori S., Fabian A.C., 1997, MNRAS, 292, L33
\bibitem[]{} Ettori S., Fabian A.C., White, D.A., 1998, MNRAS, 300, 837
\bibitem[]{} Evrard A.E., Metzler C.A., Navarro J.F., 1996, ApJ, 469,
494
\bibitem[]{} Evrard A.E., 1997, MNRAS, 292, 289
\bibitem[]{} Fabian A.C., 1991, MNRAS, 253, 29p
\bibitem[]{} Fabian A.C., 1994, ARAA, 32, 277
\bibitem[]{} Freedman W.L., Mould J.R., Kennicutt Jr R.C., Madore 
B.F., 1998, astro-ph/9801080
\bibitem[]{} Fukazawa Y., Makishima K., Tamura T., Ezawa H., Xu H.,
Ikebe Y., Kikuchi K., Ohashi T., 1998, PASJ, 50, 187
\bibitem[]{} Fukugita M., Hogan C.J., Peebles P.J.E., 1998, ApJ, 503, 518
\bibitem[]{} Girardi M., Escalera E., Fadda D., Giuricin G., 
Mardirossian F., Mezzetti M., 1997, ApJ, 482, 41
\bibitem[]{} Henriksen M.J., Markevitch M., 1996, ApJL, 466, L79
\bibitem[]{} Hogan C.J., 1998, Space Science Reviews, 84 (1/2), 127
\bibitem[]{} Hjorth J., Oukbir J., van Kampen E., 1998, MNRAS, 298, 1
\bibitem[]{} Kaastra J.S., 1992, {\it An X-Ray Spectral Code for
Optically Thin Plasmas} (Internal SRON-Leiden Report, updated version 2.0)
\bibitem[]{} Liedahl D.A., Osterheld A.L., Goldstein W.H., 1995, ApJ,
438, L115
\bibitem[]{} Loewenstein M., Mushotzky R. F, 1996, ApJL, 471, 83
\bibitem[]{} Makino N., Sasaki S., Suto Y., 1998, ApJ, 497, 555
\bibitem[]{} Malumuth E. M., Kirshner R.P., 1985, ApJ, 291, 8
\bibitem[]{} Markevitch M., Mushotzky R., Inoue H., Yamashita K., 
Furuzawa A., Tawara Y., 1996, ApJ, 456, 437
\bibitem[]{} Markevitch M., Vikhlinin A., ApJ, 1997, 491, 467
\bibitem[]{} Markevitch M., Forman W.R., Sarazin C.L., Vikhlinin 
A., ApJ, 1998, 503, 77
\bibitem[]{} Markevitch M., 1998, 504, 27
\bibitem[]{} Metzler C.A., Evrard A.E., 1998, ApJ, submitted
(astro-ph/9710324)
\bibitem[]{} Navarro J.F., Frenk C.S., White S.D.M., 1995, MNRAS,
275, 720 
\bibitem[]{} Navarro J.F., Frenk C.S., White S.D.M., 1997, ApJ, 490,
493 
\bibitem[]{} Nulsen P.E.J., Fabian A.C., 1997, MNRAS, 291, 425
\bibitem[]{} Owen F.N., Ledlow M.J., Morrison G.E., Hill J.M., 1997,
ApJL, 488, L15
\bibitem[]{} Perlmutter S. et al., 1998, Nature, 391, 51
\bibitem[]{} Press W.H., Teukolsky S.A., Vetterling W.T., Flannery
B.P., 1992, {\it Numerical Recipes}, Cambridge University Press
\bibitem[]{} Press W.H., 1996, "Unsolved Problems in Astrophysics",
Proceedings of Conference in Honor of John Bahcall, ed. J.P.
Ostriker, Princeton University Press
\bibitem[]{} Rines K., Forman W., Pen U., Jones C., Burg R., 1999, ApJ,
in press (astro-ph/9809336)
\bibitem[]{} Rottgering H.J.A., Wieringa M.H., Hunstead R.W., Ekers
R.D., 1997, MNRAS, 290, 577
\bibitem[]{} Rybicki G.B., Lightman A.P., 1979, {\it Radiative
Processes in Astrophysics}, John Wiley \& Sons, New York 
\bibitem[]{} Sarazin C.L., Wise M.W., Markevitch M.L., 1998, ApJ,
498, 606
\bibitem[]{} Sasaki S., 1996, PASJ, 48, L119
\bibitem[]{} Smail I., Ellis R.S., Dressler A., Couch W.J., Oemler 
A.Jr., Sharples R.M., Butcher H., 1997, ApJ, 479, 70 
\bibitem[]{} Snowden S.L., McCammon D., Burrows D.N., Mendenhall
J.A., 1994, ApJ, 424, 714 
\bibitem[]{} Songaila A., Wampler E.J., Cowie L.L., 1997, Nature,
385, 137
\bibitem[]{} Steigman G., Tkachev I., 1999, ApJ, in press (astro-ph/9803008)
\bibitem[]{} Stewart G.C., Fabian A.C., Jones C., Forman W., 1984, ApJ,
285, 1
\bibitem[]{} Struble M.F., Rodd H.J., 1991, ApJS, 77, 363
\bibitem[]{} Takizawa M., 1998, ApJ, 509, 579
\bibitem[]{} Thomas P.A., Fabian A.C., 1990, MNRAS, 246, 156
\bibitem[]{} Thomas P.A. et al., 1998, MNRAS, 296, 1061
\bibitem[]{} Walker T.P., Steigman G., Schramm D.N., Olive K.A.,
Kang H.S., 1991, ApJ, 376, 51
\bibitem[]{} White D.A., Fabian A.C., 1995, MNRAS, 273, 72
\bibitem[]{} White D.A., Jones C., Forman W., 1997, MNRAS, 292, 419
\bibitem[]{} White S.D.M., Frenk C.S., 1991, ApJ, 379, 52
\bibitem[]{} White S.D.M., Navarro J.F., Evrard A.E., Frenk C.S.,
1993, Nature, 366, 429
\bibitem[]{} Wu K.K.S., Fabian A.C., Nulsen P.E.J., 1998, MNRAS, 301, L20
\end{thebibliography}
\end{document}